\input epsf
\epsfverbosetrue
\documentclass[useAMS,usenatbib]{mn2e}
\usepackage{epsfig}
\title[Extragalactic Background Light]
{Extragalactic Background Light: a measurement at 400~nm using dark cloud shadow
\thanks{K.M., P.V., K.L. and Ch.L. dedicate this paper to the memory 
of  Gerhard von Appen-Schnur, friend and colleague, who deceased on 13 February 2013}
\thanks{Based on observations collected at the European Organisation for Astronomical 
Research in the Southern Hemisphere, 
under ESO programmes 072.A-0208(A), 082.A-0421(A), and 086.A-0201(A)}\\
{\LARGE II. Spectroscopic separation of the dark cloud's light, and results}}
\author[K. Mattila et al.]{K. Mattila$^{1}$\thanks{E-mail: mattila@cc.helsinki.fi}, 
P. V\"ais\"anen$^{2,3}$,
K. Lehtinen$^{1}$, 
G. von Appen-Schnur$^{4}$ 
and Ch. Leinert$^{5}$
\\
$^{1}$Department of Physics, University of Helsinki, P.O. Box 64, FI-00014 Helsinki, Finland\\
$^{2}$South African Astronomical Observatory, P.O. Box 9 Observatory, Cape Town, South Africa\\
$^{3}$Southern African Large Telescope, P.O. Box 9 Observatory, Cape Town, South Africa\\
$^{4}$Astronomisches Institut, Ruhr-Universit\"at Bochum,  Universit\"atsstrasse 150, D-44801 Bochum, Germany \\
$^{5}$Max-Planck-Institut f\"ur Astronomie, K\"onigstuhl 17, D-69117 Heidelberg, Germany}
\begin{document}
\newcommand{\ana}{A\&A~}  
\newcommand{\apj}{{ApJ~}}     
\newcommand{\mnras}{{MNRAS~}} 
\newcommand{\ea}{{et~al.~}}
\newcommand{\cgs}{$10^{-9}$\,erg\,cm$^{-2}$s$^{-1}$sr$^{-1}$\AA$^{-1}$}
\newcommand{\ang}{\AA \ }
\date{Accepted ; Received ; in original form }
 \pagerange{\pageref{firstpage}--\pageref{lastpage}} \pubyear{2002}

\maketitle

\label{firstpage}

\begin{abstract}

{In a project aimed at measuring the optical Extragalactic Background Light (EBL) 
we are using the shadow of a dark cloud. We have performed, with the ESO {VLT}/FORS,  
spectrophotometry of the surface brightness towards the high--galactic--latitude dark cloud 
Lynds~1642. A spectrum representing  the difference 
between the opaque core of the cloud and several unobscured positions 
around the cloud was presented in Paper I\, \citep{mat17a}.  
The topic of the present paper is the separation of  the scattered starlight from the dark cloud 
itself which is the only remaining foreground component in this difference.  While the scattered 
starlight spectrum has the characteristic Fraunhofer lines and the 
discontinuity at 400~nm, typical of integrated light of galaxies, the EBL spectrum is a 
smooth one without these features. As template for the scattered starlight we make use 
of the spectra at two semi-transparent positions.   The resulting  EBL intensity at 400~nm  is
$I_{\rm EBL}=2.9\pm1.1$\,\cgs\, or $11.6\pm4.4$\,nW m$^{-2}$sr$^{-1}$,  
which represents a 2.6$\sigma$ detection; the scaling uncertainty is +20\%/-16\%. 
At 520~nm we have set a 2$\sigma$ upper limit of $I_{\rm EBL} \le$4.5\,\cgs\, 
or  $\le $ 23.4~nW m$^{-2}$sr$^{-1}$ +20\%/-16\% .
Our EBL value at 400~nm is $\ga 2$ times as high as the integrated light of galaxies.
No known diffuse light sources, such as light from Milky Way halo, intra-cluster or intra-group stars
appear capable of explaining the observed EBL excess over the integrated light of galaxies.
}
                          
\end{abstract}


\section{Introduction}
Besides the Cosmic Microwave Background (CMB) also 'the other' extragalactic background radiations 
from $\gamma$-- and X--rays over ultraviolet (UV), optical, near--infrared (NIR) to far--infrared (FIR) 
radiation and long radio waves have been recognized as an important source of information 
for cosmology and galaxy formation studies and have been intensively studied both by 
ground-based and space-borne telescopes; for a review see e.g. \citet{longair95}. 
Energetically, the optical--NIR and the FIR components 
are (after the CMB) the most important ones. The source of energy for both of them
is mainly the starlight, the direct one in optical/NIR and the dust--processed one 
in the mid-- and far--IR. While already \citet{loys} and \citet{olbers} recognized the 
relevance of the night sky darkness for cosmology the optical EBL  has 
remained, paradoxically, the observationally least well determined component so far.  
This is the consequence of the smallness of the Extragalactic Background Light (EBL) 
as compared to the much brighter sky foreground components.

The conventional approach for EBL measurement has been to observe the total sky brightness
and try to measure or model all the foreground components. Then, after subtraction, what is left
over is the EBL. In this method, because of smallness of the EBL, the two large surface
brightness components, $I_ {\rm total}$ and $I_{\rm foreground}$, must be very accurately known and,
if measured by different telescopes or methods, {\em absolutely} calibrated at $< 0.5$\%
accuracy which for surface brightness measurements is hardly possible \citep{b7}.

When all or part of the observations are done from ground 
the first hindrance is the atmospheric diffuse light, {\bf ADL}, consisting of the Airglow, 
{\bf AGL}, and the tropospheric scattered light. The second large foreground is 
the Zodiacal Light, {\bf ZL}, which remains a problem with full weight even if the observations 
are carried out outside the atmosphere. A very promising approach is to make the measurements
beyond the interplanetary dust cloud. However, unless a small and accurately positioned aperture 
can be used, the light from stars in the measuring aperture enters with full weight and 
overwhelms the EBL \citep{Matsu11}. Even if all these nearby foregrounds had been successfully 
determined there remains, for an observer within the Galactic dust layer, the emission by 
interstellar gas and the Diffuse Galactic Light, {\bf DGL}, the starlight scattered by interstellar 
dust. No perfect cosmological windows, completely free of gas and dust, are known to exist. 
  Besides the acronyms EBL, ADL, AGL, DGL and ZL we will in the following use also {\bf ISRF}
for Interstellar Radiation Field, {\bf ISL} for Integrated Starlight and  {\bf IGL} for
the integrated light of galaxies. 

In the present paper, based on the results of \citet{mat17a}, hereafter Paper I, 
we present a (spectro)photometric determination of the EBL using the dark cloud shadow method 
\citep{mat76,mat90}. The method utilizes the screening effect of a dark cloud on the background light.
In this method no absolute measurements or modelling of $I_ {\rm total}$ and $I_{\rm foreground}$ are required. 
The difference of the night sky brightness in the direction of an opaque high galactic latitude 
dark cloud and a surrounding transparent area is due to two components only: (i) the EBL, 
and (ii) the  starlight that has been diffusely scattered by interstellar dust in the 
cloud. The three large foreground components, i.e. the ZL, the AGL, and the tropospheric 
scattered light, are eliminated. Also any emission by interstellar gas or dust that is
homogeneously distributed in front of the cloud is eliminated. A sketch shown as  Fig. 1 
of Paper I illustrates the principle of the method. 
The method utilizes one and the same instrument for all sky components and, therefore, 
does not require any higher-than-usual calibration accuracy. 

The main task in the method is to account for the surface brightness 
of the dark cloud. For the cloud, and even for its dark core, we have to take into account 
a substantial scattered light component. Such a component, 
although maybe smaller, is present in any direct photometric EBL measurement 
and is not an extra annoyance specific to the dark cloud method only. 
For efficient detection of the EBL with the dark cloud method it is essential that the 
dark cloud has an opaque core, $\tau > 10$, and that there are areas with good transparency 
in its immediate neighbourhood on sky. 
    
The rest of this paper is organized as follows.
In Section~2 the method of spectroscopic separation of the EBL from the 
scattered light is described. In Section~3 we present our
observational results and error estimates for the EBL intensity in the blue band, 
$\lambda\sim 400$~nm, and at~520~nm. In Section~4 we discuss our EBL result
in comparison with the integrated light of galaxies (as derived from deep galaxy
counts) and with the $\gamma$-ray attenuation measurements of blazars. 
In the final Section~5 we present our summary and conclusions.  

The Paper has two appendices containing supporting data. Appendix~A describes 
synthetic spectrum modelling of the the Solar neighbourhood ISL; it is used 
for comparision with the empirically determined scattered light spectrun 
in the cloud. Appendix~B  presents a survey of possible 
sources of diffuse light from outskirts of galaxies, low surface brightness
galaxies and intergalactic stars that could contribute substantially to the EBL

\section{Separation of the EBL and scattered light}

The observational result of Paper I consists of the surface brightness spectra at the 
opaque central position, Pos 8, and the semi-transparent positions 9, and 42 in L\,1642 relative
to the OFF positions: 
$\Delta I^{0}(\lambda)$(Pos8 -- OFF), $\Delta I^{0}(\lambda)$(Pos9 -- OFF), 
and $\Delta I^{0}(\lambda)$(Pos42 -- OFF).
These spectra represent the surface brightness difference relative to 
the mean sky brightness in the transparent surroundings of the cloud, represented by 
the OFF-positions 18, 20, 24, 25, 32, 34, and 36 (see Table 1 of Paper I). 
The results refer to the corrected spectra, i.e. outside-the-atmosphere.
In the following they will be referred to, for simplicity,  without the 
superscript '0'. The mean values for Pos9 and Pos42 will be designated
by $\Delta I(\lambda)$(Pos9/42 -- OFF) and  $\Delta I(\lambda)$(Pos9/42 -- Pos8).  

\subsection{The components of the cloud spectrum}
The following components contribute to the surface brightness difference
'dark cloud minus surrounding sky'. 

\noindent{\bf (A)} { We designate the isotropic EBL as seen by an observer outside 
the Galaxy $ I_{\rm EBL}(\lambda)$. It is attenuated by the intervening dust along 
the line-of-sight (LOS) and there is additional light} scattered from the all-sky isotropic EBL into 
the LOS;  (a) toward the comparison (OFF) fields we see almost unattenuated LOS EBL;
(b) toward the opaque cloud position, Pos8, we see almost completely attenuated LOS EBL
plus a weak scattered EBL signal; and (c) toward the translucent cloud positions, 
Pos9 and Pos42, we see moderately attenuated LOS EBL and a moderate scattered EBL signal.
{ We designate the modified EBL signal  $ I'_{\rm EBL}(\lambda)$. It is discussed 
in Section 2.1.1.}\\ 
\noindent {\bf (B)} Light scattered by dust with all-sky ISL plus DGL 
as source of illumination; (a) in the cloud; and (b) in the comparison (OFF) fields;
the spectrum is a modified copy of the ISL spectrum, i.e. a continuum with Fraunhofer lines 
and the 400~nm discontinuity;\\
\noindent{\bf (C)} Light scattered by dust into the LOS with diffuse all-sky ionized-gas emission 
as source of illumination; (a) toward the cloud positions; and (b) toward the 
comparison (OFF) fields. The spectrum is dominated by the Balmer lines. { The Galactic scattered
light, including the components under (B) and (C), is designated} 
$ I_{\rm SCA}(\lambda)$ { and will be discussed in Sections 2.1.2 and 2.1.3.}  \\
\noindent{\bf (D)} Direct line-of-sight diffuse emission by ionized gas;
(a) toward the dark cloud positions Pos8, Pos9 and Pos42; and (b) toward the comparison 
(OFF) positions; the spectrum is an emission line spectrum dominated by strong Balmer lines 
plus a weak continuum. { This component is designated} $ I_{\rm BAL}(\lambda)$ {and will
be discussed in Section 2.1.3.}

Components (C) and (D) contribute substantially only to the wavelength slots where  
the stronger Balmer lines are present.  The observed differential spectrum 
'dark cloud (ON) minus surrounding sky (OFF)' can be represented as follows:
\begin{equation}
\Delta I_{\rm obs}(\lambda) = \Delta I'_{\rm EBL}(\lambda)+{ \Delta I_{\rm SCA}(\lambda)}
+\Delta I_{\rm BAL}(\lambda)
\end{equation}
{ Notice that while} ${ \Delta I_{\rm SCA}(\lambda)} > 0$ { the differences}
$\Delta I'_{\rm EBL}(\lambda)$ and $\Delta I_{\rm BAL}(\lambda)$ { are negative }
since the EBL and the Balmer line contribution are larger at the OFF than at the ON positions.

\subsubsection{The EBL signal}
{ We consider first the idealized case with an EBL signal only and fully tranparent OFF areas.}
Then, for an optical depth of $\tau_{\rm cl}$ through the cloud,  we can write the difference  
$\Delta I'_{\rm EBL}(\lambda) = I'^{\rm ~on}_{\rm EBL}(\lambda)- I'^{\rm ~off}_{\rm EBL}(\lambda)$
in the form

\begin{eqnarray}
\lefteqn{\Delta I'_{\rm EBL}(\lambda)= I_{\rm EBL}(\lambda)[e^{-\tau_{\rm cl}}
+f_{\rm sca}(\tau_{\rm cl})] -I_{\rm EBL}(\lambda)={} } \nonumber \\
   &  {}= -I_{\rm EBL}(\lambda)[1-e^{-\tau_{\rm cl}}-f_{\rm sca}(\tau_{\rm cl})] =  {} \nonumber \\
   &  {}= -h I_{\rm EBL}(\lambda)
\end{eqnarray}
where $f_{\rm sca}(\tau_{\rm cl})$ is the fraction of scattered EBL from the dark cloud and

\begin{equation} 
h = 1-e^{-\tau_{\rm cl}}-f_{\rm sca}(\tau_{\rm cl}). 
\end{equation}

For a completely opaque cloud, $ \tau_{\rm cl}>>1$, with non-scattering dust grains (i.e. albedo $a = 0$) 
$f_{\rm sca}(\tau_{\rm cl}) = 0$ and  the correction factor is  $h = 1$. For a finite optical 
depth and the presence of scattered light its value is reduced below unity. 

\begin{table}
\caption{Calculated values of the effective blocking factor $h= 1-e^{-\tau_{\rm cl}}-f_{\rm sca}(\tau_{\rm cl})$ 
for  a spherical cloud of different optical thicknesses (diameters) $\tau_{\rm cl}$ and for isotropic
incident radiation. The errors reflect the uncertainties of the scattering parameters 
$a$ and $g$. Error estimates in parentheses are based on the values for $\tau_{\rm cl}=2-16$.}
\begin{tabular}{rl}
$\tau_{\rm cl}$& $h$\\
\hline
0.5    &0.18($\pm$0.03)\\
1      &0.32($\pm$0.06)\\
2      &0.53$\pm$0.10\\
4      &0.75$\pm$0.12\\
8      &0.90$\pm$0.09\\
16     &0.95$\pm$0.05\\
20     &0.96($\pm$0.05) \\
\hline
\end{tabular}
\end{table}

The calculation of $h$ for the 
isotropic incident EBL and for a homogeneous spherical cloud has
been presented in Appendix 1 of \citet{mat76}. Monte Carlo method was used to solve the radiative transfer
problem for different values of  $\tau_{\rm cl}$ (=cloud diameter)  and the 
scattering parameters of the grains, i.e. the albedo $a$ and asymmetry parameter $g$. Using the values 
$a=0.6\pm0.1$ and $g=0.7$ to 0.8 \citep{mat71} the values as given in Table 1 were found 
for $h$. They refer to the central part of the cloud disk, with an area of 1/10th of the disk area.
Towards the opaque 'standard position', Pos8, with $A_{\rm V}$(Pos8)$\ga15$~mag (see Section 3 and 
Table 1 of Paper I) 
the directly transmitted EBL through the cloud is $<10^{-5}$, and the scattered EBL does not 
influence the ON -- OFF difference $\Delta I_{\rm EBL}(\lambda)$ by more 
than 5\%, i.e. the cloud's blocking efficiency is $\sim95$\%. Towards the two intermediate opacity 
positions, Pos9 and Pos42, with  $A_{\rm V}\approx1$~mag (see Section 3 and Table 1 of Paper I), 
the cloud blocks $\sim30$\% of the EBL. Towards the OFF positions
with a mean extinction of $A_{\rm V}$(OFF)$\approx0.15$~mag the EBL blocking factor is 
 $\sim5$\%, i.e. $\sim95$\% of the EBL signal is transmitted.

\begin{figure*}
\vspace{0pt}
\includegraphics[width=110mm, angle=-90]{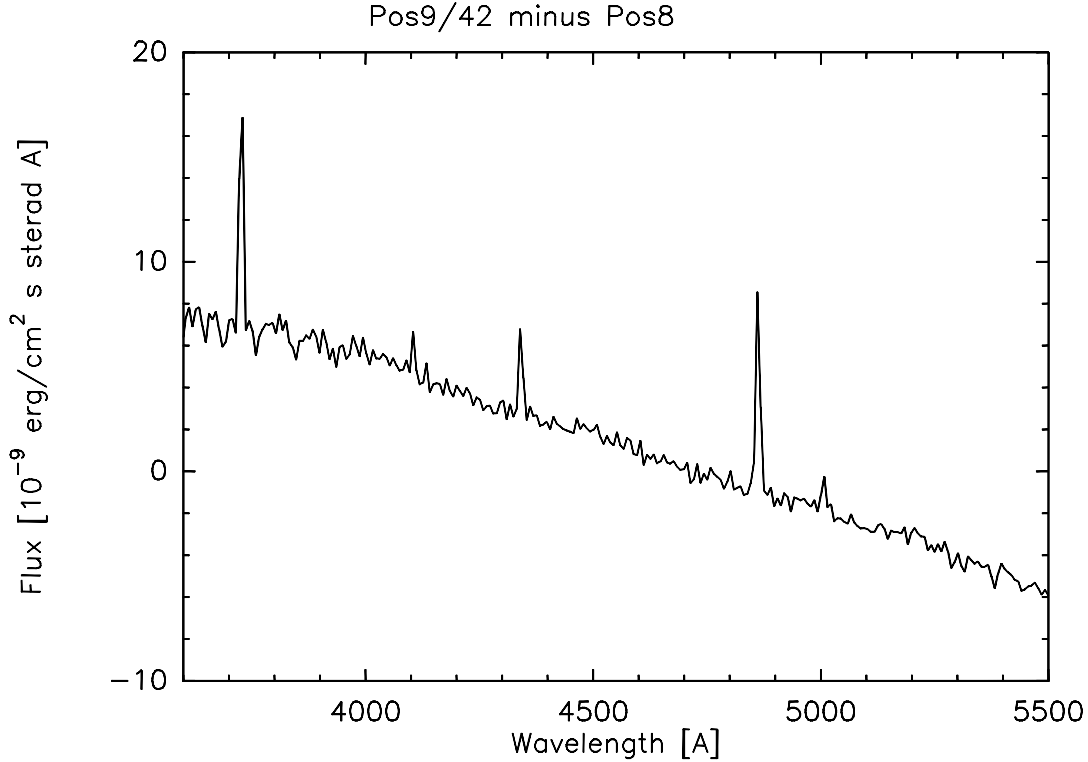}
\caption{The differential spectrum $\Delta I(\lambda)$(Pos9/42 -- Pos8). The emission lines
in this spectrum are due to the gas located {\em behind} the cloud. Besides the Balmer lines
H\,$\beta$, H\,$\gamma$ and H\,$\delta$ also the [O\,{\sc ii}] line at 372.8~nm and [O\,{\sc iii}] line at 500.7~nm
are seen. The line intensities are attenuated by extinction when transmitted through the dust
layer of $A_V \sim 1$~mag at Pos9 and Pos42. } 
\end{figure*}

\subsubsection{Scattered { Galactic} light}
The spectrum of the scattered Galactic light from the cloud can be represented as the product 
of two factors
\begin{equation}
{I^{\rm on}_{\rm SCA}(\lambda)}=\langle i_{\rm GAL}(\lambda)\rangle G_{\rm SCA}(\lambda)
\end{equation}
where $\langle i_{\rm GAL}(\lambda)\rangle$ stands for
the normalized spectrum of the impinging Galactic light; it is normalized to = 1 at a reference 
wavelength $\lambda_0$, i.e.
\begin{equation}
 \langle i_{\rm GAL}(\lambda)\rangle = \langle I_{\rm GAL}(\lambda)\rangle /\langle I_{\rm GAL}(\lambda_0)\rangle
\end{equation}
and $G_{\rm SCA}(\lambda)$ accounts for the intensity of the scattered light as well as for the 
reddening (or bluening) caused by the wavelength--dependent scattering and extinction in the cloud.  
$G_{\rm SCA}(\lambda)$ is assumed to be, over a limited wavelength range, a linear function of $\lambda$ 
 
\begin{equation}
G_{\rm SCA}(\lambda) = G_{\rm SCA}(\lambda_0)[1+grad\times(\lambda-\lambda_0)].      
\end{equation}

The intensity and spectrum of the Galactic light, as seen by a virtual observer at the 
cloud's location, is a function of the galactic coordinates, 
$I_{\rm GAL}(\lambda)=I_{\rm GAL}(\lambda,l,b)$.
The radiation impinging at the cloud's surface from different directions $l,b$ is multiply scattered by 
the dust grains in the cloud, and $\langle I_{\rm GAL}(\lambda)\rangle$
is the weighted mean of  $I_{\rm GAL}(\lambda, l,b)$ over the sky, with the direction-dependent 
weighting function being determined by the scattering properties of dust and the cloud's optical thickness.

In Appendix A a synthetic spectrum of the Galactic starlightlight 
has been calculated by using spectra of representative stars
of different spectral classes. 
The mean ISL spectrum over sky as seen by a cloud at the $|z|$-distance of 85~pc, 
representative of L\,1642, is shown in Fig.~A1 in Appendix A. 
 The { normalized} Galactic spectrum, $\langle i_{\rm GAL}(\lambda)\rangle$, can be assumed to 
be closely the same for all positions in the cloud (i.e. Pos8, 9, 42) 
{ as well as for the OFF positions as far as the strengths} 
of the Fraunhofer lines and 400~nm discontinuity are concerned. 
The absolute level and the wavelength dependence of $G_{\rm SCA}(\lambda)$ will be different
and this is accounted for by different values of  $G_{\rm SCA}(\lambda_0)$ and $grad$.

For the OFF positions, with $A_V\approx0.15$~mag (see Section 3 and Table 1 of Paper~I), 
the scattered Galactic light can be calculated using the optically thin approximation: 
\begin{equation}
I_{\rm SCA}^{\rm off}(\lambda)= C_{\rm N}\times\tau_{\rm off}(\lambda) \times  \langle i_{\rm GAL}(\lambda)\rangle
\end{equation}
The absolute level is normalized using the factor $C_{\rm N}$ so that at 555~nm 
$I_{\rm SCA}^{\rm off}(\lambda)$ = 3.34\,\cgs\,. 
This value has been determined using the intermediate band photometry at 555~nm 
combined with ISOPHOT 200 $\mu$m absolute surface photometry (see Appendix C of Paper~I).
The wavelength dependence of extinction has been adopted from \citet{car}. 
The albedo of the dust has been taken to be constant  
over the wavelength range considered in this paper; 
this is justified by the results of \citet{lau}. 
Because $\tau_{\rm off}(\lambda)$ increases toward shorter wavelengths,
$ I_{\rm SCA}^{\rm off}(\lambda)$ is bluer than the integrated { Galactic light}
represented by  $\langle i_{\rm GAL}(\lambda)\rangle$.

\subsubsection{Direct line-of-sight emission by ionized gas}

There is an all-sky diffuse light component caused by the widely distributed ionized gas. The strongest 
features in the optical spectrum are the Balmer emission lines; in addition, a weak Balmer continuum 
emission and other emission lines are present, such as  [O\,{\sc ii}]~372.8~nm, 
[O\,{\sc iii}]~500.7~nm, [N\,{\sc ii}]~654.8 and 658.3~nm, and 
[S\,{\sc ii}]~671.6 and 673.1~nm \citep{brandt,Lehtinen13}. 
The Balmer lines are present as strong absorption lines in the ISL spectrum
(see Fig. A1 of Appendix A). They could, in principle, serve as
good measures of the dark cloud's scattered ISL if there were not the 
emission lines from the ionized gas. The corrections for   H\,${\beta}$ and
 H\,${\gamma}$  are so large that these lines have to be excluded from the 
scattered light separation analysis. However, the Balmer emission lines 
at $\lambda < 400$~nm, H\,${\epsilon}$, H$_8$, H$_9$ and H$_{10}$ are much weaker 
and cause only small corrections to the ISL spectrum; their contributions can be estimated 
from H\,$\beta$, H\,$\gamma$ and H\,$\delta$ lines using the Balmer decrement valid for 
diffuse interstellar gas.

The relative Balmer line intensities have been adopted according to Table 4.2 of 
\citet{osterbrock} for Case B, low-density limit, $T_e$ = 10,000 K.
The lines are broadened to fit the observed, instrumentally broadened line profiles. 
The continuum has been adopted according to \citet{aller}, Table 4.9, $T_e$= 10,000 K.
    
When using the observed spectrum at Pos9/42 as proxy for the impinging Galactic light 
spectrum $\langle i_{\rm GAL}(\lambda)\rangle$ (see Section 2.2.3 below) 
the scattered light from not only the ISL and DGL but also from the all-sky Balmer line 
(and continuum) emission is included in it. 

\begin{table}
\caption{Intensities of spectral lines in the differential spectrum  
$\Delta I(\lambda){\rm(Pos9/42 - Pos8)}$, see Fig. 1. Unit for line areas is
 $10^{-9}$~ergs cm$^{-2}$s$^{-1}$sr$^{-1}$ and for peak intensities\,\cgs\,. The model 
fit is given for line-of-sight { optical depth of} $\tau_{1}$(H\,${\beta}$)= 0.90 { for Pos9/42}. 
In the last line is given the derived H\,$\beta$ intensity for the background emission.}
\begin{tabular}{lrrcr}
Line &\multicolumn{2}{c}{Observed line} &\multicolumn{2}{c}{Model}\\
     &\multicolumn{2}{c}{------------------------} &\multicolumn{2}{c}{-------------------}\\
     &\multicolumn{1}{c}{Area} & Peak &$ \tau_{1}$&Area\\
\hline
 H\,$\beta$             &$112.8\pm2.8$  & 9.8&0.90 &112.8 \\
 H\,$\gamma$            &$46.1\pm2.8$   & 5.1&1.03 &46.1 \\ 
 H\,$\delta$            &$18.7\pm3.2$   & 2.4&1.11 &23.4 \\ 
 $[$O\,{\sc ii}$]$~372.8~nm  &$135.0\pm6.9$  &13.2&     &    \\
 $[$O\,{\sc iii}$]$~500.7~nm & $20.7\pm2.3$  & 2.2&     &   \\
\hline
\multicolumn{3}{l}{Derived background H\,$\beta$ intensity}&\multicolumn{2}{r}{227}\\
\hline
\end{tabular}
\end{table}

Emission by the gas layer {\em in front} of the cloud cancels out, on average, in
the differential spectra $\Delta I(\lambda){\rm(Pos8 - OFF)}$ and $\Delta I(\lambda){\rm(Pos9/42 - OFF)}$.
The intensity of the { direct line-of-sight} Balmer line emission from the {\em background} gas, 
$ I_ {\rm BAL}^{\rm bg}(\lambda)$ 
can be estimated as follows: Because of the large extinction,  $ I_ {\rm BAL}^{\rm bg}(\lambda)$
is negligible toward Pos8  and only its contribution at OFF and at positions 9 and 42 
has to be considered. Because of the transmitted background light the differential spectrum  
$\Delta I(\lambda)$(Pos9/42 -- Pos8) shows the Balmer lines in emission, see Fig. 1. 
In Table 2 the line areas and peak values, as obtained from this spectrum, are given for the H\,$\beta$, 
H\,$\gamma$, H\,$\delta$, [O\,{\sc ii}]~372.8~nm and [O\,{\sc iii}]~500.7~nm lines. The peak values were 
derived from the line areas using the line width of 10.5~\AA\, as determined by the spectral resolution. 
 { We designate the optical depth through the cloud at Pos9/42 $\tau_1(\lambda)$.} Then, the background  
H\,$\beta$ intensity, $I_{\rm BAL}^{\rm bg}$(H$\beta$), can be calculated from \\
$\Delta I({\rm H}\beta)({\rm Pos9/42 - Pos8})= I_ {\rm BAL}^{\rm bg}({\rm H}\beta){\rm e}^{-\tau_1({\rm H}\beta)}$.\\
The omission of the {\em scattered} H\,$\beta$ line contribution  is justified
because the scattered light intensity at the wavelength of H\,$\beta$ is closely the same 
for Pos8 and Pos9/42 { and thus cancels out in their difference} (see Fig. 6 of Paper I). 

We have adopted for the optical depth the value $\tau_1({\rm H}\beta)$ = 0.90 which 
correctly reproduces the observed line ratio H\,$\beta$/H\,$\gamma$. It is somewhat smaller than
the unweighted mean value for Pos9 and 42, $\tau_1({\rm H}\beta)$ = $1.1\pm0.1$, but 
close to $\tau_1({\rm H}\beta)=0.94\pm0.1$ for Pos42 which dominates by its higher weight
in the spectrum $\Delta I{\rm (Pos9/42 - Pos8)}$. 
{ (For the extinction values towards Pos9 and 42 see Section 3 and Table 1 of Paper I.)}  
 
In the model fitting procedure the  intensities of the Balmer lines and Balmer continuum 
relative to H\,$\beta$ intensity are kept fixed and the whole spectrum is scaled by one and the same factor.
At $\lambda > 380$~nm the level of Balmer continuum correction 
is always $\la 0.25$\,\cgs\,. Its uncertainty is $\la10$\% which means that the correction introduces
no significant uncertainty to the EBL determination.

\subsection{Model fitting of the observed spectrum of the opaque position}

\subsubsection{Cloud geometry}

The high opacity position Pos8 towards the centre of the cloud is bracketed
by the intermediate opacity positions 9 and 42 on its northern and southern side, separated from it by
$\sim$ 10 and 7 arcmin, respectively. It is plausible that the dense core is located
at about half way inside the lower opacity halo which has an extinction 
of $A_V\approx 1$~mag. Then, a layer with  $A_V\approx0.5$~mag
would be located in front and a layer with  $A_V\approx0.5$~mag
 behind of it. Scattered light intensity from the layer in front of the core amounts
to $\sim$half of the intensity at the positions 9 and 42. The model is schematically shown in Fig.~2.
Variations on the model will be considered by placing the core at different fractional depths of
$d$ = 0, 0.3, 0.5 and 0.7 from the surface of the cloud. 

\begin{figure}
\vspace{-5mm}
\hspace{-5mm}
\includegraphics[width=95mm, angle=-0]{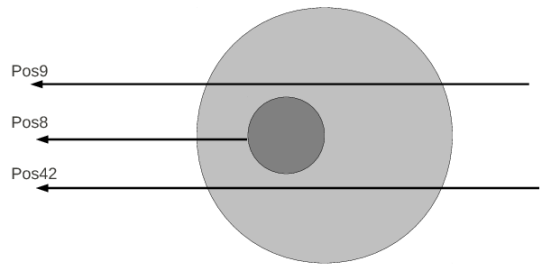}
\vspace{-60pt}
\caption{Schematic model of the cloud with opaque core embedded in a partly transparent envelope} 
\end{figure}

\subsubsection{Model fit using the synthetic ISL spectrum}

Synthetic models of the ISL spectrum 
have been calculated as described in Appendix A.
In order to test and validate such a spectrum we have compared it with the observed mean spectrum
for positions 9 and 42,  $\Delta I(\lambda){\rm(Pos9/42 - OFF)}$ (see Appendix A, Fig. A1). At these positions 
the obscuration of the EBL is small and the spectral shape is only weakly influenced by dust.
The comparison shows that while the overall fit is good for $\lambda > 400$~nm
there is a substantial discrepancy with the 400~nm step size: the 
'observed minus model' values deviate systematically upwards 
at  $\lambda < 400$~nm, i.e. the ISL model predicts a larger step at 400~nm than the 
observed one. Such a behaviour can not be explained by an EBL contribution. 
A residual EBL contribution would have an effect in the opposite direction.

Rather than trying to adjust the model parameters of the synthetic ISL spectrum we will 
in the following { use the observed Pos9/42 spectrum to derive a proxy for the  impinging 
Galactic spectrum}.

\subsubsection{Model fit using the observed scattered light spectrum}

We use the observed  ON--OFF spectrum $\Delta I(\lambda){\rm(Pos9/42 - OFF)}$ { to derive an expression for} 
the 'total power' scattered light spectrum,  $I_{\rm SCA}(\lambda)({\rm Pos9/42})$, which will be used 
as a proxy for the impinging all-sky Galactic light spectrum, $\langle i_{\rm GAL}(\lambda)\rangle$. 
\begin{eqnarray}
\lefteqn{I_{\rm SCA}(\lambda)({\rm Pos9/42})= \Delta I(\lambda)({\rm Pos9/42-OFF})+ I^{\rm off}_{\rm SCA}(\lambda)+ {} } \nonumber \\
      & {} +h_1I_{\rm EBL}(\lambda) +[1-{\rm exp}(-\tau_1(\lambda))]I^{\rm bg}_{\rm BAL}(\lambda) 
\end{eqnarray}

{ The term for the scattered Galactic light at the OFF positions, 
$I_{\rm SCA}^{\rm off}(\lambda)$, will be estimated below.}
The moderate contribution by 
the EBL signal is corrected for by the term $h_1I_{\rm EBL}(\lambda)$, where $h_ 1\approx0.32$ 
corresponding to $\tau_1\approx 1$ (see Table 1). 
{ This correction term will be included into the final component 
separation procedure to be applied to the opaque position, Pos8, spectrum.}  

The direct line-of-sight gas emission from the layer behind the cloud introduces two terms: 
(1) the transmitted light at Pos9/42, exp$(-\tau_1)I^{\rm bg}_{\rm BAL}(\lambda)$(Pos9/42), 
and (2) the emission toward the OFF positions, $I^{\rm bg}_{\rm BAL}(\lambda)$(OFF). 
While the former term is known from the analysis of the
spectrum $\Delta I(\lambda){\rm (Pos9/42 - Pos8)}$ (see Section 2.1.3 and Fig. 1) 
the latter term is not. { In equation (8) we have assumed that $I^{\rm bg}_{\rm BAL}(\lambda)$
is equal for Pos9/42 and the OFF positions.} 
Above, we have also assumed that the foreground Balmer emission is the same towards the cloud and
the OFF positions,  $I^{\rm fg}_{\rm BAL}(\lambda)({\rm Pos8})$=  $I^{\rm fg}_{\rm BAL}(\lambda)({\rm Pos9/42})$=  
$I^{\rm fg}_{\rm BAL}(\lambda)({\rm OFF})$, and has thus cancelled out in the differential spectra. 
While fitting the spectrum  $\Delta I(\lambda)({\rm Pos8-OFF})$ these approximations can be 
accounted for by adjusting the scaling of the Balmer line spectrum; it will be slightly different from the value
 found for Pos9/42, $I^{\rm bg}_{\rm BAL}(H\beta)$ = 26\,\cgs\,. { We notice that the terms $I^{\rm bg}_{\rm BAL}$
and $I^{\rm fg}_{\rm BAL}$ are important only for the wavelengths of Balmer lines.}
 
For the OFF positions, with $A_V\approx0.15$~mag, the scattered Galactic light can be represented
 using the optically thin approximation and the spectrum for positions 
9 and 42, modified by a correction for the optical depth effect, $[1 -\exp({-\tau_{1}(\lambda))}]^{-1}$:
\begin{equation}
I_{\rm SCA}^{\rm off}(\lambda)= C_{\rm N}\times \tau_{\rm off}(\lambda)
\frac{I_{\rm SCA}(\lambda)({\rm Pos9/42})}{1 -\exp(-\tau_ 1(\lambda))}
\end{equation}
Concerning the normalization  factor $C_{\rm N}$ and the wavelength dependence of the 
optical depth,  $\tau(\lambda)$, we refer to to equation (7) above. Because $I_{\rm SCA}^{\rm off}(\lambda)$ and 
$I_{\rm SCA}(\lambda)({\rm Pos9/42})$ depend on each other according to equations (8) 
and (9) their values have to be determined iteratively. In practice, 
two iterations were found to be enough. 
 
The observed spectrum $\Delta I(\lambda){\rm(Pos8-OFF)}$ to be fitted in accordance with equations (1), (2) 
and (6) is the mean of the 'Master A\&B Mean' and 'Secondary Mean' spectra (see Section 8 and Fig. 6 of Paper I). 
The spectrum is re-binned with 5-pixel
boxcar function resulting in 0.7~nm bins, closely corresponding to the physical resolution.

For the fitting procedure we used the {\sc idl}\footnote{http://www.exelisvis.com/ProductsServices/IDL.aspx} 
programme {\tt MPFITFUN}\footnote{www.physics.wisc.edu/~graigm/idl/fitting.html}.
The function {\tt MYFUNCT} used for the fitting of  $\Delta I(\lambda){\rm(Pos8 - OFF)}$ is formulated as follows:
\begin{eqnarray}
\lefteqn{{\tt MYFUNCT}= [p_0+p_1(\lambda-\lambda_0)]I_{\rm SCA}(\lambda){\rm (Pos9/42)} -I_{\rm SCA}^{\rm off}(\lambda)  {} } \nonumber \\
   &  {}-h_0I_{\rm EBL}(\lambda)-p_2I_{\rm BAL}(\lambda) 
\end{eqnarray}
The parameters $p_0$ and $p_1$ correspond to $G_{\rm SCA}(\lambda_0)$ and $grad$ in equation (6); 
for $\lambda_0$ we use 400.0~nm. For $I_{\rm SCA}(\lambda)$(Pos9/42) and  
$I_{\rm SCA}^{\rm off}(\lambda)$ we use
the spectra derived according to equations (8) and (9)  from the spectrum $\Delta I(\lambda){\rm(Pos9/42-OFF)}$.
The EBL intensity, $I_{\rm EBL}(\lambda)$ 
is assumed to be constant over each of the limited wavelength intervals
used for the fitting procedure. 
For the parameter $h_0$  corresponding to $\tau\ga15$ for Pos8 we adopt according to Table 1 the value $h_0=0.95$.
The direct line-of-sight Balmer emission will be scaled using the parameter $p_2$. 

Equation (10) corresponds to the geometric model where the opaque core 
is located on the near side of the cloud envelope, i.e. $d=0$. 
As discussed in Subsection 2.2.1 above, in a more likely geometric configuration
the core is at an intermediate depth, say $d\sim0.5$, within the cloud. 
We have run alternative model fits with this geometry in mind.
Model-wise, we can peel off the the foreground dust layer by modifying the observed spectrum for Pos8.
In the case that the core is located at a depth  $d$ within the cloud
the observed spectrum for Pos8 is replaced by \\
$\Delta I(\lambda){\rm(Pos8-OFF)}-d\times\Delta I(\lambda){\rm(Pos9/42-OFF)}$ \\
and {\tt MYFUNCT} is modified as follows:
\begin{eqnarray}
\lefteqn{{\tt MYFUNCT}=  {} } \nonumber \\
    &  {}  =[p_0+p_1(\lambda-\lambda_0))]I_{\rm SCA}(\lambda)({\rm Pos9/42})-(1-d)I_{\rm SCA}^{\rm off}(\lambda) {}  \nonumber \\
    &  {}   -(h_0-h_1d)I_{\rm EBL}(\lambda)-(1-d+de^{-\tau_1(\lambda)})p_2I_{\rm BAL}(\lambda) 
\end{eqnarray}

In the {\tt MPFITFUN} fitting run $p_0$, $p_1$ and  $I_{\rm EBL}$ are free parameters which are simultaneously 
determined by least squares fitting. The scaling factor $p_2$
for the Balmer emission is determined in a separate fitting procedure applied to the Balmer lines only, 
and its value is kept fixed in the main fitting procedure.
The quality of the fit is judged by the $\chi^2$ value that results from the fit.

\begin{figure*}
\vspace{0pt}
\includegraphics[width=110mm, angle=-90]{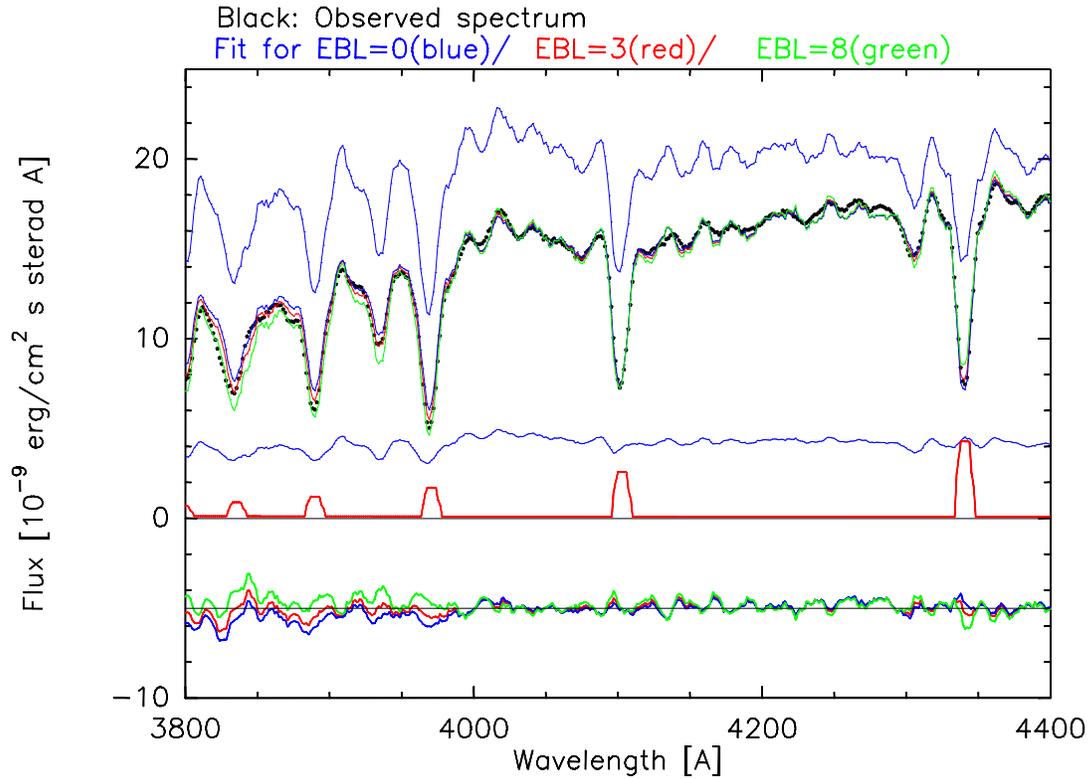}
\caption{Demonstration of the use of the 400~nm discontinuity for the determination
of the EBL. The spectrum $\Delta I(\lambda){\rm(Pos8-OFF)}$ of the opaque central position 8 
(black dots) is fitted with the Pos9/42 spectrum (topmost blue curve).  
Different values have been assumed for the EBL to demonstrate how they influence 
the goodness of the fit. The lower blue line is the mean 
scattered light OFF spectrum according to equation (9) and the red line the 
direct line-of-sight ionized gas emission line OFF spectrum. The model fits
are shown superimposed on the observed Pos 8 spectrum
for three values, $I_{\rm EBL}$ = 0 (blue), 3 (red) and 8\,\cgs\,(green).
The residuals {\em observation minus fit} are shown at the bottom
with the same colour coding. See text for details.} 
\end{figure*}

\begin{figure*}
\vspace{00pt}
\includegraphics[width=70mm, angle=-90]{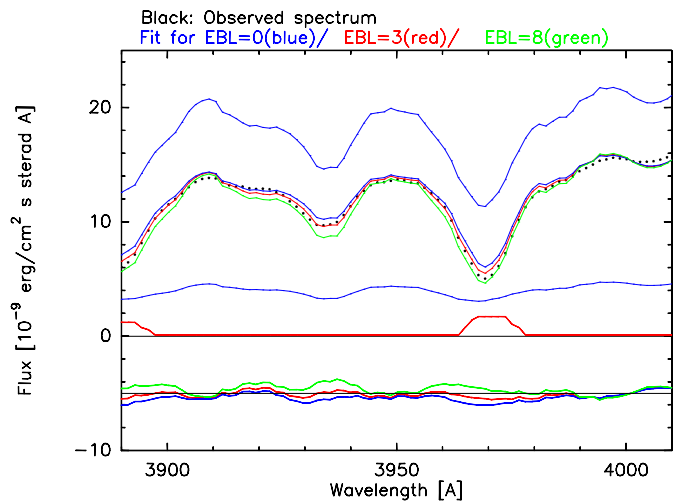}
\caption{Demonstration of the use of the Ca H and K lines for the determination
of the EBL as in Fig. 3. The 390--400~nm spectrum of the opaque position 8 (black dots) 
is fitted with the Pos9/42 spectrum (topmost blue line) for three assumed values 
of the EBL. The lower blue line is the  
scattered light spectrum and the red line the ionized hydrogen emission line spectrum 
for the OFF positions. The fits are shown superimposed on the observed spectrum for position 8
for three values of the EBL intensity, $I_{\rm EBL}$ = 0 (blue), 3 (red) and 8\,\cgs\,(green).
The residuals {\em observation minus fit} are shown as the bottom-most three curves 
with the same colour coding and
with zero level shifted by -5\,\cgs\,. See text for details.} 
\end{figure*}

\section{Results from {\tt MPFITFUN} fitting}

Promising spectral features for the separation of the scattered light 
are the 400~nm discontinuity, the strong Fraunhofer lines H and K of Ca\,{\sc ii} at 397 and 393~nm, 
and the G band at 430~nm. To a lesser extent the Mg{\sc i}+MgH band at 517~nm and the Fe line 
at 527~nm can be useful for setting at least an upper limit to $I_{\rm EBL}$. 
Balmer lines H\,$\beta$, H\,$\gamma$
and H\,$\delta$ would appear as good candidates because they are strong in the ISL spectrum
(see Fig. A1 of Appendix A). However, because they are present in the direct line-of-sight 
ionized gas emission as well, they do not enable an estimation of the scattered light. 
Nevertheless, one can use them to estimate the contamination 
effect which  H\,$\epsilon$ has on the  Ca\,{\sc ii} H line, and  H$_8$,  H$_9$ and H$_{10}$ on 
the overall mean intensity level at 380--390~nm.

\subsection{Results from different wavelength slots}
We  first demonstrate qualitatively that the spectroscopic separation method
is capable of reaching the required sensitivity level of $\la1$\,\cgs\,.
In Fig. 3 we demonstrate the use of the 400~nm discontinuity to the determination
of the EBL. The 380--440~nm spectrum of the opaque central position 8 
(black dots) is shown together  with the Pos9/42 spectrum
according to equation (8). The lower blue curve at $\sim 4$\,\cgs\,is the  scattered light spectrum 
for the OFF positions (equation (9)). The red line shows the correction for the 
direct line-of-sight Balmer emission lines. 

Model fits are shown superimposed on the observed spectrum for  $I_{\rm EBL}$ = 0 (blue), 
3\,\cgs\, (red), and 8\,\cgs\,(green), respectively.
The residuals, {\em observation minus fit}, are shown as the bottom-most three curves. 
 Their deviations from zero level
at $\lambda<400$~nm reflect the influence
of the 400~nm  step size. For $I_{\rm EBL}$ = 0 the step size of the model fit
is smaller while for  $I_{\rm EBL}$ = 8\,\cgs\, it is larger than the observed one.
For an intermediate value of  $I_{\rm EBL}$ = 3\,\cgs\, a good agreement is reached.

In Fig. 4  we demonstrate the use of the H and K lines to the determination
of the EBL. The 390--400~nm spectrum of the opaque position 8 (black dots) 
is again fitted with the Pos9/42 spectrum (blue line) for three values 
of the EBL. The fitted spectra are deciphered in the same way as in Fig. 3.
Comparison of the fitted line depths with the observed ones indicates
that for  $I_{\rm EBL}$ = 0 the model predicts too shallow lines while for 
$I_{\rm EBL}$ = 8\,\cgs\, they are deeper than the observed ones. Again, the
intermediate value of  $I_{\rm EBL}$ = 3\,\cgs\, gives the best fit. 

In Table 3 we give  $I_{\rm EBL}$ values and their statistical errors, $\sigma_{\rm fit}$, 
obtained from {\tt MPFITFUN} fitting for seven cases. The wavelength slots 
used are given in the second column. 
The third column gives the relevant spectral features that
are facilitating the separation of the scattered light from the EBL.
The EBL values and their $1\sigma$ statistical errors 
or $2\sigma$ upper limits are given in the fifth and sixth column. 
Systematic errors follow in column 7.

The most sensitive probe is provided by the case where both the H and K 
lines and the 400~nm discontinuity are included in the fit. 
The first result is for the 50~nm wide range, $\lambda$ = 381--432.5~nm,
which includes also the Balmer lines H\,$\delta$ to H$_9$.
The Balmer line scaling factor is included as a free parameter in the fitting.
Its value is thereby well determined mainly by the H\,$\delta$ line which is not blended with
other spectral features. 

Although all the Balmer lines at  $\lambda < 400$~nm are substantially weaker than  
H\,$\delta$ there may still be a residual effect, especially because of the 
H\,$\epsilon$ line at 397~nm that is blended with the  Ca\,{\sc ii} H line. 
As the second case we have, therefore, made a fit to the same wavelength range, $\lambda$=381--432.5~nm,
but excluding all Balmer lines and thereby also the H line of  Ca\,{\sc ii}.

In the third case even the K line has been excluded and the result is thus based
mainly on the 400~nm jump only. In the fourth case we use a narrow window 
including the H and K lines only, while in the fifth case only the K line is included.  

The error estimates, $\sigma_{\rm fit}$, as given by {\tt MPFITFUN}, 
are seen to increase while the wavelength window is narrowed down. 
The first four cases provide significant $I_{\rm EBL}$ values ($\ga 2\sigma$ detections) 
in agreement with each other.
In the fifth case the error is too large to allow an EBL detection.

A fit to the G band alone (Case 6) can be used to set
an upper limit of $I_{\rm EBL} \le 5.0$\,\cgs\, $(2\sigma)$. The region 510-530~nm includes 
two relatively strong Fraunhofer lines, MgH+Mg{\sc i} at 517 and Fe at 527~nm. The fitting allows
an upper limit to be set to the EBL  at 520~nm, 
$I_{\rm EBL} \le 4.5$\,\cgs\, ($2\sigma$) (Case 7). 

\begin{table*}
\begin{minipage}{175mm}
\caption{EBL values from {\tt MPFITFUN} fitting  utilizing different spectral features of the scattered ISL.
Statistical errors are given as resulting from the pixel-to-pixel noise of the spectra (see Section 3.1), 
the zero level uncertainty of the  $\Delta I(\lambda){\rm(Pos9/42 - Pos8)}$ spectrum (Section 3.2), 
and the Balmer line correction (Section 3.3). 
Their quadratic sum is $\sigma_{\rm tot}$. The upper limits given for cases 6 and 7 include also the systematic 
error caused by the zero level uncertainty of the Pos8-OFF spectrum (Section 3.2).}
\begin{tabular}{clllcll}
No. &Wavelength range [nm] &{Spectral features}&Remarks &$I_{EBL}$\footnote{ $I_{EBL}$ and errors are units of\,\cgs\, 
except for  $\sigma_{\rm cal}$ and $\sigma_{\rm model}$ which are in per cent of the $I_{EBL}$ value}&
Statistical errors$^{\it a}$ & Systematic err$^{\it a}$\\
    &                     &used for fit& &            &       
$\sigma_{\rm fit}$,$\sigma_{\rm zero}$,$\sigma_{\rm Bal}$, $\sigma_{\rm tot}$,  $\sigma_{\rm cal}$&$\sigma_{\rm zero}$, $\sigma_{\rm model}$ \\
(1)&\multicolumn{1}{c}{(2)}&\multicolumn{1}{c}{(3)}&\multicolumn{1}{c}{(4)}&(5)&\multicolumn{1}{c}{(6)}&\multicolumn{1}{c}{(7)} \\
\hline
\hline
1 & 381-432.5 & 400~nm jump,    &H\,$\delta$,H\,$\epsilon$,H\,$\zeta$,H$_9$&3.20& 0.81\ 0.70\  0.3\ \  1.11 \ \ 9\% & 0.13\ \ $^{+18\%}_{-13\%}$ \\ 
  &           & G,  Ca\,{\sc ii} H+K & included                            &    &                       & \\ 
\hline
2 & 381-382.5,385-387.5,     & 400~nm jump,   &Balmer lines \&&2.54 &0.94\ 0.70\  \ \   - \ \   1.17 \ \ 9\% & 0.13\ \ $^{+18\%}_{-13\%}$  \\ 
  &390-396,398-408,412-432.5 & G,  Ca\,{\sc ii} K      & Ca\,{\sc ii} H excluded    &     &                       & \\ 
\hline
3 & 381-382.5,385-387.5,     & 400~nm jump, &Balmer lines \& & 2.54\ &1.16\ 0.70\ \ \   -\ \ \  1.35\  \ \ 9\% & 0.13\ \ $^{+18\%}_{-13\%}$ \\ 
  &390-392.5,398-408,412-432.5 & G band     & Ca\,{\sc ii} H+K excl   &       &                         & \\  
\hline 
4 & 390-399           &  Ca\,{\sc ii} H+K &H\,$\epsilon$ included            & 3.58  &1.42\  0.70\   0.5\ \ 1.66  \ \ 9\% & 0.13\ \ $^{+18\%}_{-13\%}$ \\
\hline
5 & 390-396           &  Ca\,{\sc ii} K &    Ca\,{\sc ii} K only                    & 1.76  &1.86\ 0.70\ \  \    -\ \ \  1.99 \ \ 9\% & 0.13\ \ $^{+18\%}_{-13\%}$      \\
\hline
6 & 425-432.5,436-445 & G band  &G band only &\multicolumn{2}{l}{$\le 5.0(2\sigma)$}    & \\
  &                   &         &            &0.84  & 1.91\ 0.70\ \  \   -\ \ \  2.03   \ \ 9\% & 0.13\ \ $^{+18\%}_{-13\%}$       \\ 
\hline
7 & 510-519, 522-530  & Mg517,Fe527  &520~nm Airglow  &\multicolumn{2}{l}{$\le 4.5(2\sigma)$ }  & \\
  &                   &              &line excluded &-1.80            &2.40\  1.20\ \ \    -\ \ \   2.70 \ \ 9\%  & 0.84 \ \ $^{+18\%}_{-13\%}$\\
\hline                         
\end{tabular}
\end{minipage}
\end{table*}

\begin{figure*}
\vspace{0pt}
\includegraphics[width=130mm, angle=-90]{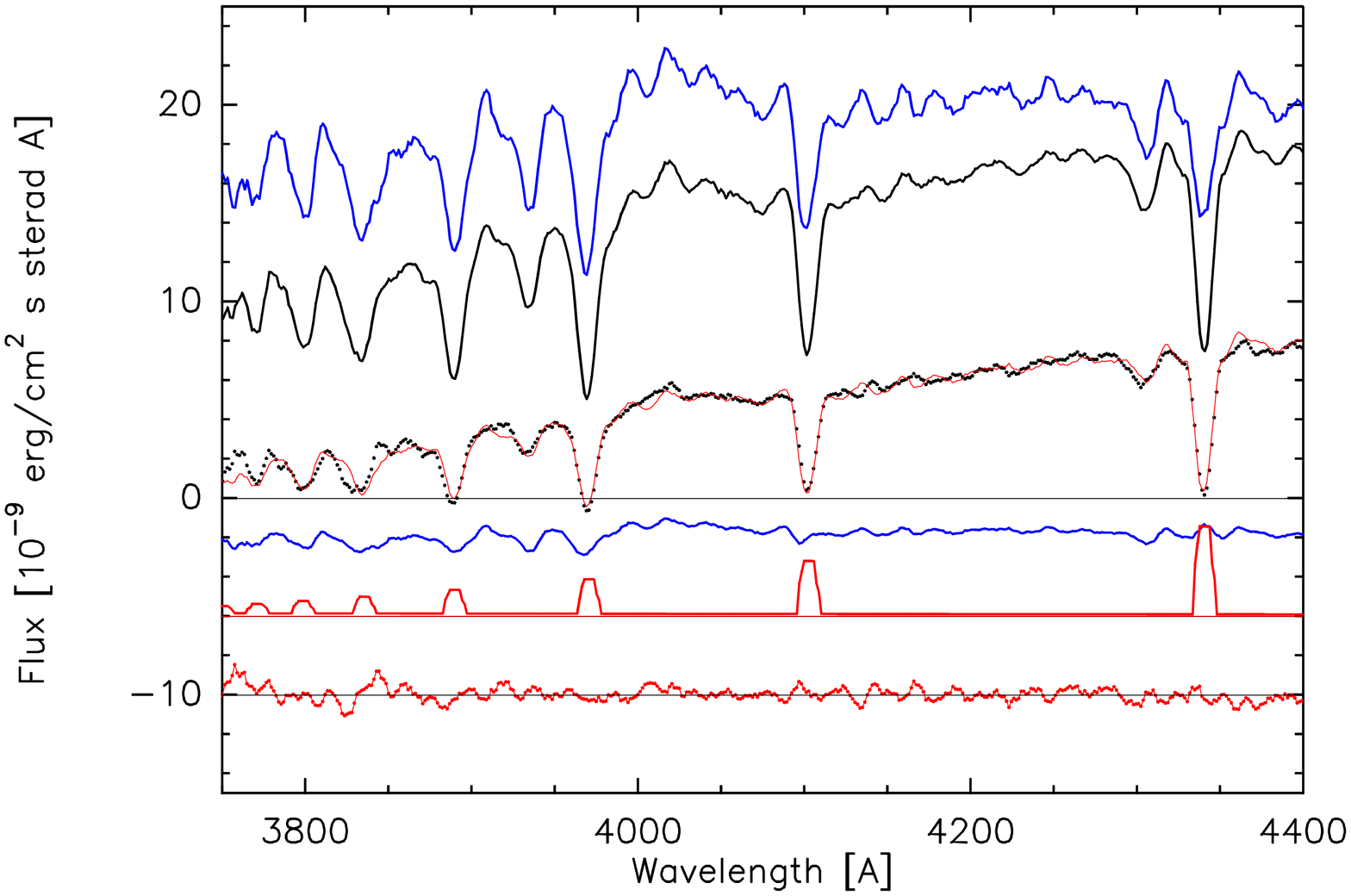}
\caption{Demonstration that the derived EBL value is not sensitive to the assumed location
of the opaque core within the cloud envelope. The 
observed spectrum for position 8 is shown as upper black line 
and the 'peeled off' spectrum after subtraction of
$0.5\times$ the Pos9/42 spectrum as black dots. 
Pos9/42 spectrum is  the upper blue line.
The lower blue line is the scattered light and the red line the gas emission spectrum 
for the OFF positions (both shifted by --6 units). The modelled spectrum for $I_{\rm EBL}$ = 3\,\cgs\,
is shown as red line superimposed on the 'peeled off' dense core spectrum.
The residuals {\em observation minus fit} are shown at bottom (shifted by --10 units). 
See text for details.}
\end{figure*}

\subsection{Errors caused by zero level uncertainties of the observed spectra}

The observational errors of the spectra $\Delta I(\lambda){\rm(Pos8-OFF)}$, 
$\Delta I(\lambda){\rm(Pos9/42-Pos8)}$ and  $\Delta I(\lambda){\rm(Pos9/42-OFF)}$ can be separated 
into three categories: (1) the pixel-to-pixel statistical errors; (2) the error of the zero level, which
for  $\Delta I(\lambda){\rm(Pos8-OFF)}$ is partly statistical and partly caused by the uncertainties of the 
differential ADL and ZL corrections; 
and (3) the  calibration errors. 

The effect of the pixel-to-pixel statistical errors
is taken into account by the {\tt MPFIFUN} fitting procedure.
The effect of calibration errors is straight forward and will be added to the errors
resulting from the modelling (see Section 3.4 below).

It has been found in Section 8.1 of Paper I that the zero point error of $\Delta I(\lambda){\rm(Pos8-OFF)}$ 
is $\pm$0.5\,\cgs\,. Because the spectrum $\Delta I(\lambda){\rm (Pos9/42-OFF)}$ was calculated as the sum of 
$\Delta I(\lambda){\rm(Pos9/42-Pos8)}$ and $\Delta I(\lambda){\rm(Pos8-OFF)}$  
it is subject to the same zero level error as  $\Delta I(\lambda){\rm(Pos8-OFF)}$. In addition,
it has the zero point error inherent in the spectrum  $\Delta I(\lambda){\rm(Pos9/42-Pos8)}$. This
error was found to be $\sim\pm0.47$\,\cgs\, at $\lambda \approx 400$~nm; 
and $\sim \pm0.63$\,\cgs\,  at $\lambda \approx 520$~nm (see Section 8.1 and Table 3 of Paper I).
For the spectrum $\Delta I(\lambda){\rm(Pos9/42-Pos8)}$, because the positions 8, 9 and 42 have small angular 
separations, the  ADL and ZL corrections are unimportant and the error is purely statistical.

In order to investigate the effects of these zero point errors we have run the MPFITFUN fitting procedure\\
(1) for the three cases where a zero point correction of -0.5, 0.0, and +0.5\,\cgs\, was applied to both 
$I(\lambda){\rm(Pos8-OFF)}$ and $I(\lambda){\rm(Pos9/42-OFF)}$; and 
(2) for each of these three cases we have applied, in addition, a correction to the spectrum  
$I(\lambda){\rm(Pos9/42-OFF)}$ of -0.47, 0.0, and +0.47\,\cgs\,
for  $\lambda$ = 381--433~nm, or -0.63, 0.0, and +0.63\,\cgs\, for  $\lambda \approx 520$~nm. 

The former zero point corrections change the EBL value by $\pm0.13$\,\cgs\,. 
Because this error is partly caused by uncertainties in the
differential ADL and ZL corrections we conservatively assign it as systematic error.
It is given as the first item ($\sigma_{\rm zero}$) in column (7).   
The latter zero level uncertainty causes a statistical error. It amounts to $\sigma_{\rm zero}=\pm0.70$  
and  $\pm1.20$\,\cgs\,  at $\lambda$ = 381--433~nm and 520~nm, respectively. 
It is given as the second item in the sixth column of Table 3.

\subsection{Errors caused by modelling uncertainties}

In addition to the observationally based errors 
there are errors caused by uncertainties in the modelling. 
These are the uncertainties of $I^{\rm off}_{\rm SCA}(\lambda)$, 
the scattered light intensity at the OFF positions;
the scaling of $I_{\rm BAL}(\lambda)$, the direct line-of-sight Balmer line intensity; 
the uncertainties of the effective blocking factors $h_0$ and $h_1$; 
and the influence of the wavelength-dependence of $I_{\rm EBL}$.

\subsubsection{OFF-position scattered light intensity}
We have varied $I^{\rm off}_{\rm SCA}(\lambda)$, given by equation (9),
by scaling it up and down by 30\%. This corresponds to the extinction uncertainty of 
$\pm0.05$mag at the OFF positions. This scaling had no effect on the derived EBL value.
This, at first sight, surprising result can be understood because the spectral features
in the OFF spectrum are the same as in the Pos9/42 spectrum used for the fitting.
Thus, the scaling of the OFF spectrum does not change the strengths (equivalent widths)
of the spectral features in the  Pos9/42 spectrum.
The spectral separation method is sensitive to the different absorption feature 
strengths of the Pos8 and Pos9/42 spectra, and not to their overall scaling difference.

\subsubsection{Balmer line intensities}
The Balmer line scaling factor $p_2$ (see equations (10) and (11)) was determined using the H\,$\delta$ line.
While  H\,$\delta$ itself was included only in the first EBL fit as given in Table 3 it was used
to give the scaling for  H\,$\epsilon$ included in the EBL determination for 
the wavelength slot 390--399~nm (fourth case in Table 3). The statistical 
uncertainty of  $p_2$, as determined from the fitting of H\,$\delta$ line   
was found to be $\pm10$\%. This causes to the EBL estimate an error of $\pm0.5$\,\cgs\,
as given as the fourth value in column (6) of Table 3. 

\subsubsection{Case of the embedded dense core}
Fig. 5 illustrates the case where an envelope layer has been 'peeled off' in front of
the opaque core (see model in Section 2.2.1 and Fig. 2). 
The core has been assumed to be half way between
the front and back surface of the envelope, i.e. $d=0.5$. The spectrum to be fitted in this case is\\ 
$\Delta I(\lambda){\rm(Pos8-OFF)}-0.5\times\Delta I(\lambda){\rm(Pos9/42-OFF)}$,\\ 
and equation (11) with $d = 0.5$ will be used in {\tt MPFITFUN} instead of equation (10). A fit to the spectrum
in the interval 380--432.5~nm with  $I_{\rm EBL}= 3$\,\cgs\, is shown as red line superimposed 
on the 'peeled off' spectrum shown as black dots.

Somewhat surprisingly, the result for  $I_{\rm EBL}$ is the same as for 
the case  $d = 0$. Also for the other values,  $d = 0.3$ and 0.7, the same is true.
This result can be understood because the spectrum of the surface layer,
removed in front of the core, has the same Pos9/42 spectrum that is used
in the fitting procedure.
This outcome demonstrates that our fitting results for the EBL do not
depend on the geometrical model, i.e. on the assumed location of the 
opaque core within the cloud envelope.

\subsubsection{Effect of the cloud's blocking factor}
We have varied the the effective blocking factors from
their adopted values of $h_0=0.95$ and $h_1=0.32$
by $\pm0.05$ and $\pm0.06$, i.e. by their uncertainties as listed in Table 1 for
$\tau_{cl}\ge 16$ and $\tau_{cl}=1$.
The maximum deviations of the $I_{\rm EBL}$ value
of +18\% and -13\% resulted for the parameter pairs $h_0=1.0$, $h_1=0.26$ and $h_0=0.90$, $h_1=0.38$, 
respectively. 

\subsubsection{Effect of the EBL spectral shape}

We have finally tested the effect of a non-constant $I_{\rm EBL}(\lambda)$ spectrum.
We assume a linear dependence with wavelength of the form   
$I_{\rm EBL}(\lambda) = I_{\rm EBL}({\rm 400~nm})[1 + C(\lambda - {\rm 400~nm})]$ and
with a large gradient, $C = +0.5$ and -0.5 per 100~nm, respectively. 
For these two extreme gradients the resulting  $I_{\rm EBL}$  values for the second case
in Table 3 differed by $\le \pm3 \%$. The effect on the other  $I_{\rm EBL}(\lambda)$
values in Table 3, covering the same or smaller wavelength range, is equal or smaller
than this. 

\subsection{The EBL results at 400 and 520~nm}

In Table 3 the first two EBL estimates are based on the same wavelength slot,
$\lambda$ = 381--432.5~nm, one with and the other without the Balmer lines. 
The two  $I_{\rm EBL}$ values are not independent. 
The good  agreement between them demonstrates, however, that our correction 
for the Balmer line contamination is reasonably good. 
The two strongest spectral features of the scattered Galactic light spectrum in this
wavelength range are the 400~nm jump and the H and K lines of  Ca\,{\sc ii}. 
The third case in Table 3 excludes, besides all Balmer
lines and  H, also the K line. The result is thus based essentially on the
400~nm jump only. 

In the fourth case we have limited the wavelength range to a narrow window
covering the H and K lines only.  The result is thus independent 
of the 400~nm discontinuity. Using a window that includes the K line only
(Case 5) the fitting error becomes too large to allow a useful $I_{\rm EBL}$ 
estimate.

All $I_{\rm EBL}$ values as derived for cases 1--4 agree within a fraction of their 
$1\sigma_{\rm tot}$ statistical errors. The first two values, based on the largest
number of spectral elements, have the smallest statistical errors. Their unweighted mean value
can thus be considered a good choice  in this
wavelength range: $I_{\rm EBL} = 2.87\pm1.1$\,\cgs\,. Because the two values are
not independent the error of their mean is not reduced below the individual errors. 
Another estimate with comparable precision is the mean of the third and fourth $I_{\rm EBL}$ values
which are independent.
Their weighted mean and its standard error is $I_{\rm EBL} = 2.95\pm1.0$\,\cgs\,. 
As weights we have used the  $\sigma_{\rm tot}^{-2}$ values. 

The two mean values (1 \& 2) and (3 \& 4)
are, somewhat fortuitously, almost equal and have almost the same standard error. 
We adopt their mean value as our final choice in the  wavelength region $\lambda$ = 380--432.5~nm: 
$I_{\rm EBL} = 2.9\pm1.0$\,\cgs\,.

In Section 3.2 we saw that an error of $\pm$0.13\,\cgs\, 
was caused by the zero level uncertainty of $\Delta I(\lambda){\rm(Pos8-OFF)}$.
If we conservatively assume that most of it is systematic
it has to be added linearly to the statistical error resulting in: $I_{\rm EBL} = 2.9\pm1.1$\,\cgs\,.

The multiplicative error consists of the spectrophotometric calibration error of $\pm9\%$
(see Section 8.2 of Paper I), 
the uncertainty caused by the blocking factor of +18\%/-13\%,
and an error of maximally $\pm3\%$ caused by possible wavelength dependence of  
$I_{\rm EBL}(\lambda)$.
Quadratic addition results in the total multiplicative error of +20\%/-16\%. 
It increases or decreases the  $I_{\rm EBL}$ 
value and its statistical error limits in the same way and, therefore, does not
influence the statistical significance level ($2.6\sigma$) of our EBL detection.

Our result for the wavelength range $\lambda = 381 - 432.5$~nm can thus be written 
in the form:\\ $I_{\rm EBL} = 2.9\pm1.1$\,\cgs\, +20\%/-16\%, or\\ 
$11.6\pm4.4$\,nW m$^{-2}$sr$^{-1}$ +20\%/-16\%.

The two upper limits for $I_{\rm EBL}$, at $\lambda \sim 430$~nm and at $\lambda \sim 520$~nm,
given in Table 3 as cases 6 and 7, are based on their statistical error estimates, 
$\sigma_{\rm tot}$, complemented by the partly systematic error caused by the
zero level uncertainty of  $\Delta I(\lambda){\rm(Pos8-OFF)}$.
In Case 6 ($\lambda \sim 430$~nm) the error caused 
by the zero uncertainty is again $\pm 0.13$\,\cgs\,. In Case 7 
 ($\lambda \sim 520$~nm) it is $\pm 0.84$\,\cgs\,. The multiplicative error is the same 
as above. 

We thus end up with the following final $2\sigma$ upper limits:\\
 $I_{\rm EBL} \le$ $5.0$\,\cgs\, +20\%/-16\%, or\\ 
$\le 20.0$ \,nW m$^{-2}$sr$^{-1}$ +20\%/-16\% at 430~nm, and\\  
$I_{\rm EBL} \le $ $(4.5$\,\cgs\, +20\%/-16\%, or \\
$ \le 23.4$ \,nW m$^{-2}$sr$^{-1}$ +20\%/-16\%  at 520~nm.

\begin{figure*}
\vspace{0pt}
\includegraphics[width=120mm, angle=-90]{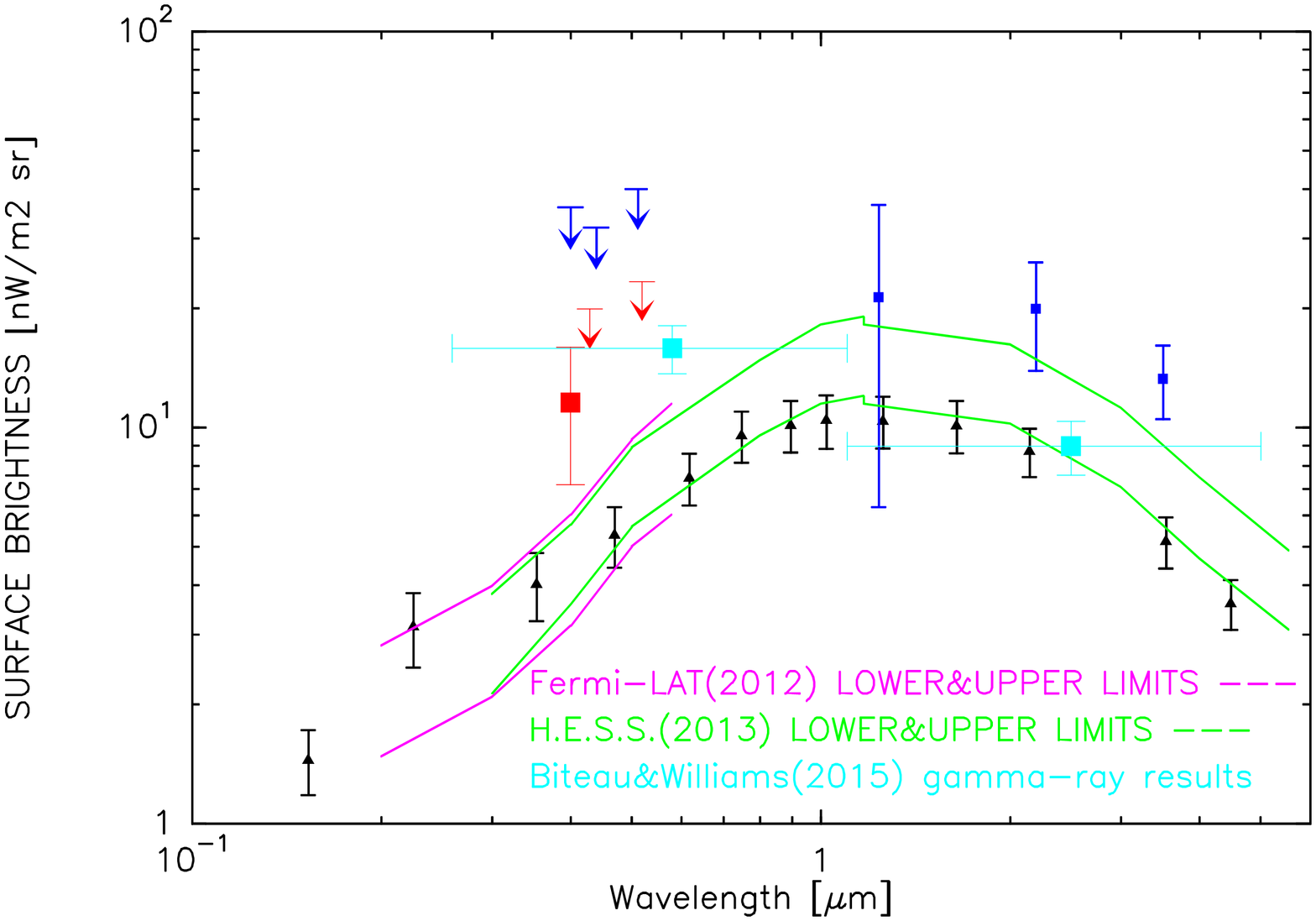}
\caption{Selection of EBL measurements, upper and lower limits. 
{\em Integrated Galaxy Light (IGL)}: 0.15--4.5~$\mu$m, \citet{driver}, black triangles. 
{\em Photometric EBL measurements and upper limits}: present paper: red square with 
$1\sigma$ error bars at 400~nm and  $2\sigma$ upper limits at 430 and 520~nm; 
dark blue squares and arrows: 
0.40~$\mu$m, \citet{mat90};  0.44~$\mu$m, \citet{Tol}; 0.5115~$\mu$m, \citet{Dube}, 
the latter three according to re-discussion in \citet{Leinert}; 
1.25, 2.2 and 3.6~$\mu$m, \citet{Levenson}. 
{\em Results from gamma-ray attenuation:} 
upper and lower limits between 0.20--0.58~$\mu$m from {\em Fermi}-LAT \citep{Ackermann}, 
magenta lines; between 0.30--5.5~$\mu$m from H.E.S.S. \citep{hess}, green lines; 
results of \citet{biteau} are shown as light blue crosses.} 
\end{figure*}

\section{Discussion}
\subsection{Contribution by resolved galaxies}

Part of our measured EBL signal is caused by resolved galaxies in the slit area. 
The slit positions were chosen to avoid stars and galaxies visible 
on the DSS2 blue plates. This corresponds to $B\approx22$ mag. In order to measure the contribution
of galaxies fainter than this limit we have secured for each one of our fields as 
listed in Table 2 of Paper I a direct $\sim$ 6.8 x 6.8~arcmin $B$ band image using {VLT} with FORS1 or FORS2,
see Table 4. For Positions 18, 20, 24, 34 and 36, with integration time 
of  $t_{\rm int}$ = 600~s,  the limiting magnitude was $\sim$26.5~mag, 
and for Pos 25 and 32 with $t_{\rm int}$ = 90~s it was $\sim$25.5~mag.
Accurate positioning of the slit relative to the direct image was secured 
by taking  immediately before each spectrum exposure a 10~s acquisition image
after which the telescope pointing was kept fixed.

The images were bias-subtracted and flat-fielded and zero point and extinction corrections
were applied using the ESO Quality Control nightly zero points and extinction coefficients 
for FORS1\footnote{http://www.eso.org/observing/dfo/quality/FORS1/qc/\\ \~/zeropoints/zeropoints.html} 
and FORS2\footnote{http://www.eso.org/observing/dfo/quality/FORS2/qc/\\ \~/zeropoints/zeropoints.html}.
Kron magnitudes of all objects were determined using the SExtractor software \citep{bertin} from within 
the Starlink {\sc gaia} v4.4.6 package. 

The objects in the slit areas were visually inspected and most appeared to be galaxies although at 
$m>23$~mag the star/galaxy separation remains uncertain. According to \citet{Windhorst} the fraction of stars
is $\la 10\%$ and  $\la 1\%$ at $g\approx23$ and 26~mag, respectively (see Appendix B1).
In column 8 of Table 4 we give the contribution  by all objects in the slit area in units of\,\cgs\,. 
Only a few per cent of this will be due to stars and we interpret it as
the EBL contribution by the galaxies in the covered magnitude interval, given in column 7.
The magnitude range varies from field to field with the bright and faint end values between 
21.9--24.0 and 24.3--26.8~mag, respectively. The galaxy contributions for the 'Master' spectra areas
vary from 0.40 to 0.58\,\cgs\,, with a mean of 0.51\,\cgs\,. For the 'Secondary' spectra fields
the corresponding numbers are 0.29 to 1.11, with a mean of 0.54\,\cgs\,. 

These numbers agree closely with the EBL contribution  of 0.49\,\cgs\, for the magnitude interval 
of 22 to 27~mag, as estimated 
from galaxy count data in the HST F435W band by \citet{driver} (their Table 3).
We conclude that our total EBL intensity of $I_{EBL} = 2.9\pm1.1$\,\cgs\, includes a contribution of
$\sim$0.5\,\cgs\, by galaxies in the magnitude interval of $\sim$22--26.5~mag.

 \begin{table*}
 \centering
 \begin{minipage}{150mm}
  \caption{Contribution of resolved galaxies in the 2 arcsec slit
at the OFF positions. The columns give position name (1) and indicate whether chip 1, 2 or both were used 
for the spectrum (2). Then follows the date of the spectrum (3). 
In columns (4) to (6) the date, integration time (in seconds) and filter 
are given for the images. Columns (7) and (8) give the magnitude range 
of galaxies and their contribution to the EBL in units of \cgs\,.}
  \begin{tabular}{@{}lccccccc@{}}

\hline
Posit- &Chip& \multicolumn{2}{c}{Date} & Int. & Filter & Magnitude   & EBL \\
ion    &    & Spectrum & Deep image    & time &        & range       & contribution\\
(1)    & (2)& (3)      & (4)           & (5)  & (6)    & (7)         & (8) \\
\hline 
\multicolumn{8}{l}{Master spectra}\\
POS18& 1   & 2003-10-20 & 2011-10-02& 600 & B\_HIGH& 24.0-26.8 &0.50\\
POS20&1\&2 & 2010-12-14 & 2010-12-14& 600 & B\_HIGH& 22.4-26.1 &0.57 \\
POS24& 1   & 2003-10-20 & 2009-01-29&600  &B\_HIGH & 23.3-26.3 & 0.40\\            
POS24&1\&2 & 2004-09-18 & 2009-01-29&600  &B\_HIGH & 23.2-26.3 &0.58  \\
 Mean&     &            &           &     &        &           &0.51 \\ 
\hline
\multicolumn{8}{l}{Secondary spectra} \\
POS18 &1\&2&   2011-10-02& 2011-10-02 & 600 & B\_HIGH&24.0-26.8 & 0.42\\
POS24 & 1\&2&  2004-02-18& 2009-01-29 & 600 &B\_HIGH &23.7-26.2 & 0.44\\
POS25  &1\&2&  2004-01-25& 2004-01-25 & \ 90  &B\_BESS &22.3-24.3 & 0.44\\
POS32 &1\&2&   2004-01-25& 2004-01-25 & \ 90  &B\_BESS &21.9-24.4 & 0.74\\
POS34a &2&     2004-09-16& 2009-01-30 & 600 & B\_HIGH &22.7-26.5 &1.11\\
POS34b &1\&2&  2004-01-24& 2009-01-30 & 600 & B\_HIGH &23.6-25.6 &0.29\\
POS36 &1\&2&   2003-11-24& 2011-10-27 & 600 & B\_HIGH &23.0-26.0 &0.36\\ 
 Mean&     &            &           &     &           &          &0.54 \\ 
\hline
\end{tabular}
\end{minipage}
\end{table*}

 \subsection{Comparison with the integrated light of galaxies (IGL) }

We show in Fig.~6 a selection from UV to NIR (0.1 - 5~$\mu$m) of direct EBL measurements and 
upper and lower limits (colour symbols and lines), as well as lower limits set by
the integrated galaxy light (IGL) from galaxy counts as derived by \citet{driver} (black symbols). 
The EBL values from the present paper are shown as the red solid square with $1\sigma$ 
error bars at 400~nm and as $2\sigma$ upper limits at 430 and 520~nm.
For references see the figure legend. 

When comparing our total EBL value at 400~nm with the IGL from galaxy counts one should take into account
the magnitude limits of galaxies included in each case. As stated above, our EBL values exclude galaxies
brighter than $B\approx22$~mag whereas the extrapolated eIGL values of \citet{driver} 
cover the the whole magnitude range, including extrapolations below $m\approx 10$ and    
above $m\approx$28--30~mag in optical bands from $u$ to $h$. 
Using their Table 3 one can estimate that galaxies with $g\le$22~mag contribute $\sim$0.52\,\cgs\,  of the total IGL of  
$1.14\pm0.2$\,\cgs\,. Thus the IGL contribution to be compared with our EBL measurement is 
$I_{\rm IGL}(g\ge22^{\rm m})=0.62\pm0.2$\,\cgs\,. On the other hand, if we include the bright-galaxy contribution
for $g \le 22$~mag, to our measured EBL value it will be increased to $I_{\rm EBL}$ = 3.5\,\cgs\,.

\subsubsection{Cosmic variance and galaxy counts}
We are able to check directly whether the L\,1642 EBL fields are abnormal in terms of IGL using 
the direct imaging of the OFF fields described in Section 4.1.  The photometric 
catalogues of the fields were cleaned of stars with the help of the SExtractor CLASS parameter (though the 
star contribution is significant only at $B<22$, see \citealt{Windhorst}) and also of obvious artifacts and edge effects.  
The resulting $B$-band galaxy counts per deg$^2$ and 0.5~mag bin, averaged over the 7 fields, are shown in Fig.~7.
The counts from the shallower images, the Pos25 and Pos32 fields, are included 
only for $B \le 24$~mag.  The counts are not completeness corrected.  For comparison we also show recent 
wide-field galaxy counts adopted from a large compilation of multi-wavelength data by \citet{driver}. 
The points shown are $g$-band counts from GAMA and COSMOS/G10 surveys as described therein, we have 
merely shifted them using an average $B-g$ = 0.6~mag conversion \citep{Fukugita}.  The galaxy counts 
in our fields down to $B\sim26$ mag thus appear totally normal and consistent with wide field galaxy counts.  

The error bars in Fig.~7 are the standard deviations of counts between the fields, the nominal Poissonian 
errors within a field are in fact much smaller.  The scatter reflects the cosmic variance.  For the size 
(0.0126 deg$^2$) of our OFF fields we expect the cosmic variance  
to be of the order of 15-20\% \citep{driver10}, consistently with our results.

Finally, we note that there is a galaxy cluster, Abell~496, just north of the L\,1642 area. 
One of our OFF positions, Pos18, is at the projected distance of 36.5 arcmin from the 
cD galaxy at the cluster centre. This corresponds to 1.63 Mpc at the cluster distance of 153.5 Mpc.
While Pos18 shows one of the two highest galaxy counts of the 7 fields examined, 
the difference is well within the scatter, the cosmic variance. An upper limit to the the diffuse 
intra-cluster light contribution from Abell~496 can be estimated from the survey by \citet{gonzalez}
of 24 clusters having a bright central galaxy with diffuse cD envelope. In this sample the highest diffuse 
light signal at large radii (up to $\sim400$ kpc) was for Abell~2984. Extrapolated to the projected 
distance of 1.63 Mpc and assuming a $B-I$ colour index of $\ge2$ mag it gives a $B$ band upper 
limit of $\le 0.01$\,\cgs. Photoelectric photometry of intra-cluster light in Abell~496 by \citet{schnur} 
gave an upper limit of $\la2$\,\cgs at the projected radial distance of 270-600 kpc.  
Adopting the radial gradient of Abell~2984 this extrapolates for 
1.63 Mpc to  $\la$0.2\,\cgs. 
We conclude that Abell~496 does not have a measurable effect on our analysis.  
 The contribution by Abell~496 can also be considered as a natural ingredient in the 
cosmic variance; in an area of $\sim4\degr\times4\degr$ as covered by our OFF positions there 
normally exists, by chance, one major Abell cluster.

\begin{figure}
\vspace{-0pt}
\includegraphics[width=90mm, angle=0]{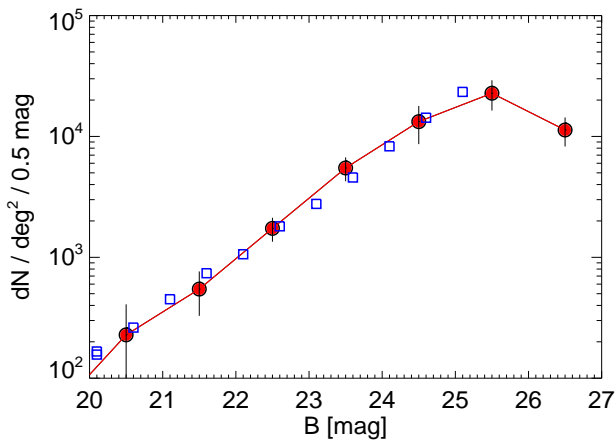}
\caption{The red circles show the average galaxy number counts in the OFF fields, the error bars depict 
the standard deviation between counts of the 7 separate fields.  The counts are not completeness corrected, 
and the incompleteness is obvious at $B > 26$ mag.  The blue squares show the g-band GAMA and G10 galaxy 
counts adopted from  \citet{driver}, using an average $B = g+0.6$~mag conversion.}
\end{figure}

\begin{figure}
\vspace{0pt}
\includegraphics[width=60mm, angle=-90]{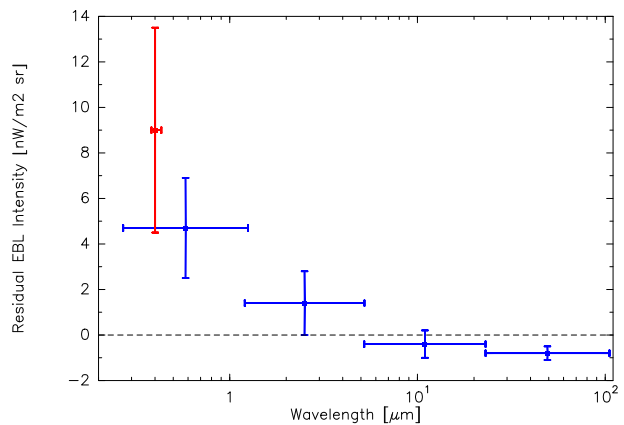}
\caption{EBL residual values after subtraction of the Integrated Galaxy Light (IGL) contribution as 
given in \citet{driver}. The four data points in blue have been adopted from Fig.~4 of \citet{biteau} 
and are  based on gamma-ray absorption measurements. The point in red is from the present study. } 
\end{figure}

 \subsection{Comparison with EBL from gamma--ray attenuation measurements}

The intergalactic radiation density has been derived from gamma-ray attenuation measurements of blazars
by ground based Cerenkov telescopes, e.g. H.E.S.S., MAGIC and VERITAS, sensitive in the 0.2--20 TeV (VHE) range, 
as well as by the Large Area Telescope (LAT), aboard the {\em Fermi} satellite, sensitive in the 1--500 GeV (HE) 
range. Results based on extensive data sets from  {\em Fermi}-LAT \citep{Ackermann} and H.E.S.S. \citep{hess} are 
shown in  Fig.~6 as upper and lower boundaries to the allowed range of EBL values. Similar results have been
published by the MAGIC collaboration \citep{ahnen}. {\em Fermi}-LAT probes with high sensitivity the EBL photon energy 
range from UV to optical ($\lambda \approx$ 200--600~nm) while the ground-based VHE telescopes
cover the wider range from optical to FIR, $\lambda \approx$ 0.3--100~$\mu$m. 

As can be seen from Fig.~6 the above mentioned gamma-ray attenuation results 
agree with  the IGL values from galaxy counts but they do allow a modestly higher
(by up to $\la 70\%$) EBL intensity as well. 
The good agreement of spectral shapes is, however, a direct result of the 
underlying assumption of an IGL--like template spectrum: the EBL spectrum template comes 
from a model that is obtained by summing up empirically--constrained galaxy populations spectra,
see e.g. \citet{frances} and \citet{Dom}. The observed gamma-ray optical depths are fit to optical depths 
predicted by EBL spectral models, with a normalisation factor as the only free parameter.
Thus, the derived EBL spectrum will closely follow the spectrum of the IGL.  

The interaction of gamma-rays vs. optical and infrared EBL photons has a very broad wavelength response function, 
with the kernel covering roughly a decade of the EBL spectrum \citep{biteau}. Therefore, the intensity 
maximum at $\sim 1$ $\mu$m dominates the normalisation factor over the whole UV--NIR range 0.2--5~$\mu$m
and it will be difficult, if not impossible, to detect deviations from the IGL-like spectral shape using the 
'model template approach'. 

Recently, \citet{biteau} have taken a more general approach. They abstain from 
an {\em a priory} assumption on constraining the EBL spectrum with the IGL--like spectrum shape. 
Using gamma-ray observations only
they were able to derive a 'free-standing' EBL spectrum which covered the  0.2--100~$\mu$m range by 
coarse binning, with four spectral elements with widths of $\sim$\, half a decade. Two of these spectral 
elements fall into the wavelength range displayed in Fig.~6 and they are shown as light blue crosses.
The horizontal bars indicate the wavelength coverages of the bins, 0.26--1.2~$\mu$m and 
1.2--5.2~$\mu$m, respectively. 

 \citet{biteau} have compared their EBL values with appropriately weighted IGL mean values
in the four bins. While a good individual and overall agreement was found for the three bins at  
 1.2--5.2~$\mu$m,  5.2--23~$\mu$m, and  23--103~$\mu$m, the bin at  0.26--1.2~$\mu$m showed an 
excess of $4.7\pm2.2$\,nW m$^{-2}$sr$^{-1}$. This excess over the IGL value, as well as over the 
{\em Fermi}--LAT and H.E.S.S. EBL values, is also well demonstrated by Fig.~6.

To compare with our EBL measurement we note that while the gamma-ray based EBL value of 
$15.9\pm2.2$\,nW m$^{-2}$sr$^{-1}$ for the bin
 0.26--1.2~$\mu$m corresponds to the total EBL, our EBL measurement at 0.4~$\mu$m does 
not include the contribution of the bright galaxies with  $g \le$22~mag. Including that
contribution, our EBL value increases from $11.6\pm4.4$\,nW m$^{-2}$sr$^{-1}$, as shown in  
Fig.~6, to $13.6\pm4.5$\,nW m$^{-2}$sr$^{-1}$ which is in good agreement with the gamma--ray value
of  \citet{biteau}. 
Furthermore, our total EBL value is in excess over the IGL value of \citet{driver} at 0.4~$\mu$m 
by $9.0\pm4.5$\,nW m$^{-2}$sr$^{-1}$. 

We show in Fig.~8, in accordance to Fig.~4 of \citet{biteau},
our EBL excess value together with their four gamma-ray based residual values.
It can be seen that our excess value, while agreeing within the error limits with the
0.26--1.2~$\mu$m gamma--ray--based value, suggests an even larger excess at 0.4~$\mu$m. 

In conclusion, our result that the EBL intensity at 400~nm exceeds the IGL from galaxy counts
is supported by the \citet{biteau} EBL value, determined by the independent method
of gamma--ray absorption. 

\subsection{Optical EBL in context of other wavelengths}

{ 
The galaxies, intergalactic stars or any other light sources that give rise to the optical EBL
will contribute to the background light also in the adjacent UV and NIR wavelength bands. We will briefly
review the recent results in these bands and their relevance for the optical EBL.

In the near  ($\lambda\approx150$~nm) and far ultraviolet ($\lambda\approx230$~nm) some of the 
large foreground sky components, e.g. the ZL,  that plague  
optical EBL measurements are absent or much reduced. However, other difficulties 
appear, see e.g. \citet{murthy09,murthy14a,murthy14b} for a review.    
The {\em Galaxy Evolution Explorer (GALEX)} \citep{martin} has recently provided a comprehensive survey 
of the near (NUV) and far ultraviolet (FUV) background, covering $\sim75$ per cent of the sky. Among the 
other surveys, \citet{schimi} have covered a large fraction of sky at 174~nm in the {\em NUVIEWS} rocket
experiment. The minimum sky brightness seen by {\em GALEX} toward the North and South Galactic Poles was
$\sim300 - 400$ (FUV) and  $\sim600$ photons~cm$^{-2}$s$^{-1}$sr$^{-1}$\AA$^{-1}$ (NUV) (see \citealt{hamden} and 
\citealt{murthy17}). While these values still contain airglow as 'likely the dominant contributor' 
\citep{hamden} they can be used as upper limits to the EBL: $I_{\rm EBL}(\lambda)\la6 -8$ 
and $\la$12 nW~m$^{-2}$sr$^{-1}$ in the FUV and NUV band, respectively. These values 
are by a factor of $\sim4$ higher than the corresponding IGL values as displayed in Fig.~6. 

Using a model for the distribution of dust and stellar light sources and adopting likely 
scattering parameters for the 
grains \citet{murthy17} estimated the scattered light from dust to be $\sim$1.2 - 1.8 (FUV) and   
$\sim$1 nW~m$^{-2}$sr$^{-1}$ (NUV). \citet{schimi} estimated from their {\em NUVIEWS} data an EBL 
intensity of $4\pm2$ nW~m$^{-2}$sr$^{-1}$ at 174~nm. We conclude that the EBL estimates at FUV and NUV
do allow an EBL contribution by at least 2 times as large as the IGL derived from galaxy counts, and
are in this respect compatible with the EBL excess at 400~nm found in the present paper. 

\citet{henry} have presented a different interpretation for the $\sim 300$  
photons~cm$^{-2}$s$^{-1}$sr$^{-1}$\AA$^{-1}$ isotropic component of the {\em GALEX} FUV sky: 
they argue that most of it is not due to EBL but is 'of unknown (but Galactic) origin'.
  
The diffuse near infrared (NIR) background sky brightness has been studied by several groups using
the {\em COBE} Diffuse Infrared Background Experiment (DIRBE) data in combination with the 2MASS star catalogue 
(see \citealt{dwek05,Levenson, sano15} for reviews), the {\em AKARI} InfraRed Camera \citep{tsumura13}, 
and the {\em IRTS} Near Infrared Spectrometer \citep{matsu05, matsu15}. Most of these results are consistent,
within their large error bars, with  the values as shown in Fig.~6 at 1.25, 2.2 and 3.6~$\mu$m 
according to the DIRBE - 2MASS analysis of \citet{Levenson}. The consensus,
shared also by the TeV gamma-ray absorption results (Section 4.4 and Figs.6 and 8),
appears to be that the NIR EBL does not exceed the IGL by more than a factor of 2. 
 
 \citet{matsu05}, however, using their {\em IRTS} data have announced and \citet{matsu15} repeated the 
claim for detection of an excess emission of up to ~6 times as large as the IGL at $\lambda = 1 - 2 ~\mu$m.
A large excess has also been found by \citet{sano15} at  $1.25 ~\mu$m  and by \citet{matsuura} 
at  $1 - 1.7 ~\mu$m.  It has been poined out by e.g. 
\citet{dwek05} and \citet{mat06} that a likely explanation for the excess found in \citet{matsu05} is 
the insufficient subtraction of Zodiacal Light. Besides the ZL also the DGL introduces substantial 
uncertainty, especially at the shorter NIR wavelengths, into all these results. 

Because of the persistent problems caused by the ZL 
the attention has turned towards the NIR sky fluctuatuations 
 (see e.g. \citealt{kash05,kash12}, \citealt{matsu11}, \citealt{cooray}). Recently \citet{Zem}, 
using the Imager instrument of the {\em Cosmic Infrared Background Experiment (CIBER)}, 
found in the auto-correlation spectra of the fluctuations at 1.1 and 1.6 $\mu$m
evidence for an excess power  at angular scales $l= \pi/\theta< 5000$ (angular separations 
$\theta > 4.3$~arcmin). When interpreted as diffuse light from galaxy clusters and groups 
(jointly called intra-halo light, IHL), the modeling \citep{cooray} of this fluctuation excess 
lead them to conclude that the IHL contributes to the {\em mean} EBL (i.e. the 'dc component') at  
1.1 and 1.6 $\mu$m 0.7 and 1.3 times as much light as the IGL from the galaxy counts. 
The interpretation has been challenged, however, by \citet{yue} and \citet{mitchell} who suggest that most 
of these observed large-angular-scale fluctuations are due to the foreground DGL.

We conclude that direct photometric measures, both in the UV (100 -- 300~nm) and NIR ($1 - 5~\mu$m), 
are compatible with an extra EBL component beyond the IGL from the galaxy counts of roughly
the same amount as the IGL itself. However, the the TeV gamma-ray absorption measurements, especially
according to \citet{biteau}, seem to exclude any substantial NIR excess (see Section 4.3 and Figs.~6
and 8) while they support an optical excess, such as found in the present paper. 
}

\subsection{Possible sources of background light from Milky Way halo and outside of galaxies}

A survey for light sources that could explain the observed EBL excess
over the integrated light of galaxies is presented in Appenxix B. Possible light sources 
that are {\em known} to exist include light from Milky Way halo, from the outskirts of galaxies, 
or from intergalactic stars in galaxy clusters and groups.
The contribution of such light sources may in certain objects or environments be very substantial,
e.g. the intergalactic stars contribute up to 40\% of the luminosity of some clusters or groups. 
However, none of these sources are capable of explaining a substantial increase of the overall
mean EBL beyond the IGL as derived from galaxy counts. 
Because of Lyman line and continuum absorption the redshift range of light sources contributing 
to sky brightness at $\lambda \la 450$ mm is limited to redshifts $z\la3.5$. 
Contributions by primordial objects such as population III stars or direct--collapse black 
holes are thus excluded.
Light from hypothetical decaying dark matter particles, such as axions, remains an open field.  

\section{Summary and conclusions}

This paper is based on the results of Paper~I \citep{mat17a}
where we have presented in the area the high galactic latitude dark cloud L\,1642 spectrophotometric 
surface brightness observations performed with VLT/FORS at ESO/Paranal. 
Spectra for several transparent sight-lines were observed {\em differentially} relative to the 
opaque core of the cloud.
Because the transparent and opaque
positions have been observed with exactly the same instrumental setup and in rapid time sequence no special
requirements arised for {\em absolute} calibration accuracy. This is a decisive advantage as compared with those
projects which derive the EBL as a (small) difference between two, $\sim$20 to 100 times larger, 
surface brightness signals measured by different telescopes and different techniques, requiring 
an extremely high {\em absolute} calibration accuracy, separately for each. 

The dark cloud offers an opaque screen toward which the EBL is close to zero. 
However, it produces also a foreground component, the scattered light from the dust.
This, the only, foreground contaminant is an order of magnitude weaker than
Zodiacal Light and Integrated Starlight, the main contaminants in the EBL measurement methods
of \citet{b1} and \citet{Matsu11}. We have accomplished the  
separation of the scattered light at the opaque position using the different
spectral characteristics of the scattered Galactic starlight and the EBL spectrum. As template for
the scattered starlight we have made use of the observed spectrum at semitransparent positions
in the cloud.   

The main results and conclusions of this paper can be summarized as follows:\\
(1) The EBL has been detected at 400~nm at 2.6$\sigma$ level. At 520~nm we have set a significant upper limit.\\
(2) The EBL value at 400~nm is\\ 
$2.9\pm1.1$\,\cgs\, +20\%/-16\%, or \\
$11.6\pm4.4$\,nW\,m$^{-2}$sr$^{-1}$ +20\%/-16\%
which is $\ga 2$ times as high as the IGL, even if possible contributions by galaxy halos to the IGL
are included.\\
(3) The 2$\sigma$ upper limit to EBL at 520~nm is\\ 
$ \le 4.5$\,\cgs\,  +20\%/-16\%, or  \\
$\le$ 23.4\,nW\,m$^{-2}$sr$^{-1}$  +20\%/-16\%.\\
(4) Our EBL value at 400~nm is in good agreement with the indirect measurement via 
gamma-ray attenuation of blazars,
presuming that the attenuation analysis has been performed, like in \citet{biteau}, without a restrictive
presupposed EBL spectral template form.\\  
(5) No diffuse light sources, such as light from Milky Way halo, intra-cluster or intra-group stars, or 
from decaying elementary particles appear capable of explaining the observed EBL excess over the IGL.

\section*{Acknowledgements}
This research has made use of the USNOFS Image and Catalogue Archive
 operated by the United States Naval Observatory, Flagstaff Station
(http://www.nofs.navy.mil/data/fchpix/).\\
The Digitized Sky Surveys were produced at the Space Telescope Science Institute 
under U.S. Government grant NAG W-2166. The images of these surveys are based on 
photographic data obtained using the Oschin Schmidt Telescope on Palomar Mountain 
and the UK Schmidt Telescope. The plates were processed into the present compressed 
digital form with the permission of these institutions. We thank Dr. Simon Driver
for valuable comments concerning galaxy counts and photometry { and the anonymous
referee for useful comments of general nature}. KM and KL acnowledge
the support from the Research Council for Natural Sciences and Engineering (Finland);
PV acnowledges the support from the National Research Foundation of South Africa.

\appendix

\section{Synthetic model of the Integrated Starlight}

Spectral synthesis is a common method in studies of stellar 
populations in external galaxies that are too distant to be resolved into 
individual stars (see e.g. \citealt{bruzual03}). We address the opposite problem: 
given the number densities and spatial distribution of the different types 
of stars and dust in the Solar neighbourhood what is the spectrum of the 
Integrated Starlight (ISL) in different directions of sky and for different 
vantage points of an observer off the Galactic plane.  
In addition to the ISL the Galactic surface brightness contains also
the diffusely scattered starlight, the Diffuse Galactic Light (DGL). 
Since its spectrum is a copy of the ISL spectrum it does not influence
the strengths of the spectral features (absorption lines, bands, discontinuities). 
Our calculation of the ISL spectrum between 370 and 600~nm
follows the methods as presented in \citet{mat80a,mat80b} with an update in 
\citet{Lehtinen13}. 
\\

\begin{figure*}
\vspace{0pt}
\includegraphics[width=130mm, angle=-90]{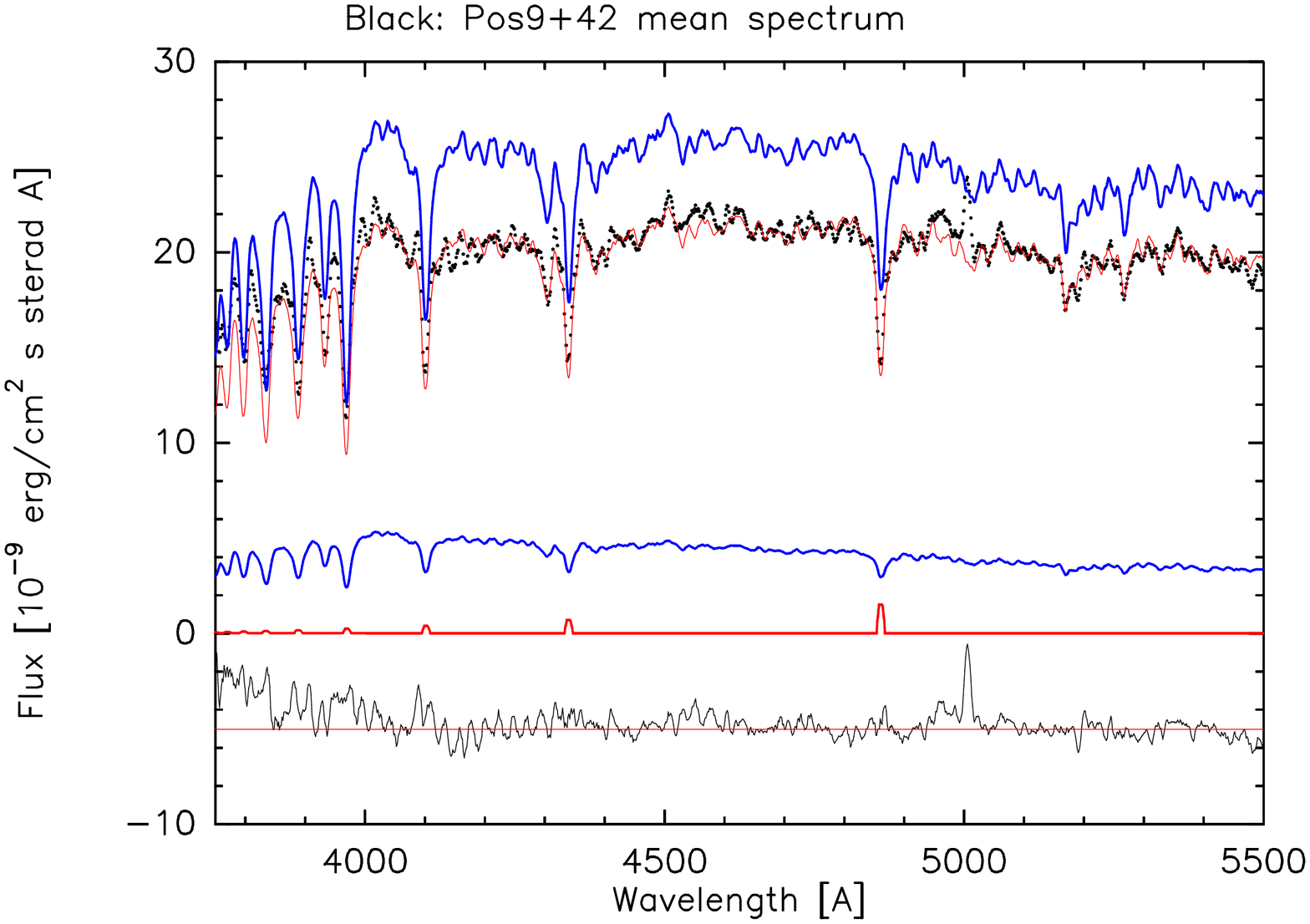}
\caption{Comparison of the mean spectrum of the intermediate-opacity positions 9 and 42 
with the synthetic ISL model spectrum (see Appendix A). The observed spectrum,  
$\Delta I(\lambda){\rm(Pos9/42 - OFF)}$, 
is represented as black dots. The ISL spectrum is shown as the uppermost blue line.
The lower blue curve is the scattered light spectrum and the red line
the ionized hydrogen emission spectrum for the OFF positions. The model
fit is shown as the red curve superimposed on the observed spectrum. 
The residuals {\em observed minus model fit} are shown as the bottom-most black curve, 
with zero level shifted by -5\,\cgs\,. Notice that the large residual at 500.7~nm
is due to the [O{\sc iii}] emission. See text for details.} 
\end{figure*}

\subsection{Galaxy model and stellar distribution parameters}

A simple model of the the Galactic structure is adopted in which  
stars and dust are distributed in plane parallel layers. The effect
of clumpy dust distribution is taken into account.
Stars have been divided into 72 spectral groups covering the different parts 
of the HR-diagram. The division  has been made according to the 
approach of \citet{flynn06} based on their analysis of the {\it Hipparcos}
data base\footnote{Drs. Flynn and Portinari kindly provided us with their results 
in detailed tabular form}. The spectral groups are combined under the following 
seven categories (see Table A1): {\it (i, ii)} Main Sequence (thin and thick disk), 
{\it (iii, iv)} Clump Stars (thin and thick disk),  {\it (v, vi)} Old Giants 
(thin and thick disk), {\it (vii)} Young Giants. For each group Table A1
gives the mean absolute magnitude, $M_V$, the number density, $D(0)$, and stellar 
emission coefficient, $j_i(0)$, in the Galactic plane, $z=0$, and the scale 
height $h_z$ for a distribution of the form $D(z) = D(0)$sech$(z/h_z)$.
Because of the limited distance range of {\it Hipparcos} its coverage for the 
supergiants was sparse. We complemented this part by using the compilation of 
\citet{Wainscoat92}.\\

\subsection{Stellar spectral library}

For the synthetic model of the 370 - 600~nm ISL spectrum we need a spectral library 
with good spectral type and wavelength coverage and a sufficient spectral resolution, 
corresponding to our observed spectra. The STELIB library 
\citep{LeBorgne03}\footnote{http://www.ast.obs-mip.fr/article181.html} matches 
well these conditions. In order to choose the best template stars from the library
for each of our stellar groups as given in Table A1 we have used besides the
spectral classes and absolute magnitudes $M_V$ also the colour indices $B-V$ 
and $V-I_c$ as selection criteria. In most cases a satisfactory match  
was possible.  The STELIB catalogue numbers are given in column 9 of Table A1. \\

\subsection{Results}

We have shown in Fig. 2 of Paper I the resulting ISL spectrum, mean over the sky, 
for an observer located at  85 pc off the Galactic plane, corresponding to the 
estimated $z$-distance of the L\,1642 cloud. In Fig. A1 we compare this ISL spectrum 
with the observed mean spectrum for positions 9 and 42,  $\Delta I(\lambda){\rm(Pos9/42 - OFF)}$. 
At these positions the obscuration of the EBL is small and the spectral shape is 
only weakly influenced by dust. The observed  spectrum is shown as black dots
and the ISL model spectrum as the uppermost blue curve. 
It has been scaled by the factor
 $G_{\rm sca}(\lambda)=0.225[1- 1.0\cdot10^{-4} {\rm~nm^{-1}}(\lambda -500 {\rm~nm})]$  
(see equation (6)) to account for a slight bluening of the Pos9/42 spectrum w.r.t. the 
ISL spectrum. The OFF spectrum, calculated according to equation (7), is shown as the lower 
blue curve. The ON -- OFF model fit, shown as the red curve, has been optimised for 
$\lambda > 400$~nm. 
The Balmer line emission in the OFF areas is shown in red with baseline at zero. 
It can be seen that while the overall fit is good for $\lambda > 400$~nm there is a 
substantial discrepancy with the 400~nm step size: this results in {\em observed minus model} 
values, shown as black line, deviating systematically upwards at $\lambda < 400$~nm, 
i.e. the ISL model predicts a larger step at 400~nm than the observed one. 
Such a behaviour cannot be explained by an EBL contribution either: 
a residual EBL contribution would have an effect in the opposite direction.

Many of the ISL spectral structures, even at low levels, are recognized also in the scattered 
light spectrum. Notice, however, the emission line at 500.7~nm in the scattered light spectrum
which is not present in the ISL; it originates from the all-sky ionized gas emission of [O{\sc iii}].

\newpage

\begin{table*}
 \centering
 \caption{Data for space distribution parameters of stars and spectral library 
(STELIB) stars used for the ISL synthetic spectrum. Column 3 gives the mean and columns 4-5 
the range of absolute magnitude, $M_V$, of the spectral group; column 6 gives the number density, 
$D(0)$ and column 8 the stellar emission coefficient, $j_i(0)$, in units of  $M_V=0$~mag stars\,pc$^{-3}$ , 
in the Galactic plane, $z=0$,; the scale height $h_z$ for a distribution of the form 
$D(z) = D(0)$sech$(z/h_z)$ is given in  column 7; the last column gives the number in the STELIB
library of the spectrum adopted as template for each spectral group. {\em The full table is available in the 
online version}.}
 \begin{tabular}{llrrrlrll}
\hline
\hline
No.& Sp class & $M_V$& \multicolumn{2}{c}{ $M_V$ range}&$D_0$&$h_z$&$j_0$               &STELIB\\  
   &          &      &       &                         &     &    &                    &star No. \\
   &       & mag  & \multicolumn{2}{c}{mag}         &stars\,pc$^{-3}$&pc& $M_V=0$~mag&  \\
   &          &      &       &                         &     &    &  stars\,pc$^{-3}$      & \\                                              
\hline
\multicolumn{9}{l}{Main sequence, thin disk} \\
 1& B0 V     & -4.0 & -4.20& -3.25 & 0.46E-06 & 56. & 0.13E-04 & 23 \\
 2& B2 V     & -2.45& -3.20& -1.85 & 0.34E-05 & 56. & 0.32E-04 & 21 \\
 3& B5 V     & -1.2 & -1.80& -0.75 & 0.13E-04 & 56. & 0.38E-04 & 94 \\
 4& B8 V     & -0.25& -0.70 & 0.20 & 0.44E-04 & 57. & 0.53E-04& 124\\ 
 5& A0 V      & 0.65 & 0.25 & 0.95 & 0.92E-04 & 68. & 0.52E-04& 163\\ 
 6& A2 V      & 1.3  & 1.00 & 1.60 & 0.15E-03 & 82. & 0.44E-04& 152\\
 7& A3-5 V    & 1.8  & 1.65 & 1.95 & 0.11E-03 & 92. & 0.21E-04& 149\\
...& ...      &...   &...   &...   &...      &...   &...      &...\\
72 & M3-4 I-II& -5.79& -5.8 & -5.8 & 0.13E-07& 56.  & 0.27E-05& 7 \\
\hline
\end{tabular}
\end{table*}

\newpage 

. 

\newpage

\section{Possible sources of EBL beyond galaxy counts: Milky Way halo, outskirts of galaxies,
low surface brightness galaxies, intergalactic stars, and decaying dark matter particles}

\subsection{Unresolved starlight and light from the Milky Way halo}
Sky background light as seen by an observer inside the Galaxy contains contributions from unresolved stars,
diffuse emission from gas, and scattered light from dust.

The  light from unresolved stars in the $B$ band  can be estimated by using the deep {\em Hubble Space 
Telescope (HST)} Advanced Camera for Surveys (ACS) starcounts 
in the F435W filter band ($\lambda_{c} = 429.7$~nm) as presented  by 
\citet{Windhorst} for the GOODS S field. At $m$(F435W) = 20 mag the number of stars is  $\sim$equal to the
galaxy numbers but, because of the shallow slope of the star counts $d(\log N(m))/dm \approx 0.065$,
it drops at $m$(F435W)= 27 mag to  $\sim$1\% of the galaxy counts. Using this slope to extrapolate 
the star counts beyond 27 mag we find that the integrated starlight for $m$(F435W)$\ge 25$ mag 
amounts to  $I_{\rm ISL}\approx0.0034$\,\cgs\,. The GOODS S field at $(l,b)$ = 224\degr, -54\degr is relatively 
close to the L\,1642 field at $(l,b)$ = 210\degr, -37\degr. Taking into account the different line-of-sight lengths 
through the Galactic (roughly plane parallel) star layer we estimate that  $I_{\rm ISL}$($B\ge$25 mag) 
$\approx0.0045$\,\cgs\, towards L\,1642 and can be neglected in the further discussion.

Extended far ultraviolet ($\lambda \approx 150 -250$~nm) halos have been observed around several edge-on 
late-type galaxies out to 5--10 kpc from the mid-plane (e.g. NGC\,891, NGC\,5907; see \citealt{seon,hodges}). 
They have been ascribed to thick dust disks that scatter light from stellar disks of the galaxies. 
The existence of substantial amounts of dust in 
the Milky Way halo at, say  $|z|> 2 $~kpc, is not known. However, any such dust would be illuminated by an 
ISRF that has closely the same absorption line spectrum as the scattered light of the L\,1642 cloud. 
Therefore, it will be largely eliminated by our spectral separation method as described in Section 2.

Gas emission from Milky Way halo is expected to be almost entirely in the form of line emission, mostly the hydrogen
Balmer lines, plus a very weak continuum. They have been taken into account in the analysis as described
in Section 2.1.3.

\subsection{Light from the outskirts of galaxies}

Not all light of a galaxy is captured by fixed aperture or Kron standard photometry.
The light loss from the outskirts of galaxies, i.e. galaxy wings, may be only partially compensated for by 
corrections towards 'total' magnitudes and this may lead to underestimates of the IGL.
 \citet{Totani01} studied its influence via photometric modelling with reasonable assumptions on galaxy 
wings and other correction factors. They found that their 'best guess' IGL from all galaxies in the Universe 
was up to 80\% higher than the IGL from a simple integration of observed galaxy counts. 

In a detailed analysis of the {\em Hubble Space Telescope (HST)} Deep Field (HDF) and their own EBL field 
 \citet{b1,b2} derived galaxy magnitudes using SExtractor Kron-type photometry with different Kron 
parameters $k=r/r_{iso}$ (their isophotal magnitudes varied from 24.7 to 25.8 ST mag arcsec$^{-2}$.) 
They found that at least 20\% of the flux of the faintest galaxies (i.e. within 4.5 mag of the detection limit)
was contributed by galaxy wings at $r > 1.4r_{iso}$. Furthermore, using a method called ``ensemble photometry''
they estimated that the true flux from $V>23$ AB mag galaxies in the HDF can be almost twice as much
as that recovered by standard photometric methods.
\citet{Benitez} analysed the faint galaxy population in two Early Release Science (ERS) fields of the 
{\em HST} ACS. They confirmed the claim of \citet{ b1,b2} of an up to 50\% loss of
light of the faintest galaxies. 
 
The galaxy photometry used by \citet{driver} for their IGL values in the optical bands is, 
at the faint end, based on the {\em HST}/ACS ERS and the 
{Hubble Ultra Deep Field} (HUDF) observations which are deeper than the 
HDF data used by \citet{b1,b2} and are, therefore, less vulnerable to effects of galaxy wings.
While both the ACS ERS \citep{Windhorst} and the HUDF photometry made use of a larger Kron factor, 
$k= 2.5$, than \citet{b1,b2} the HUDF photometry was also otherwise designed to take better
care of the galaxy-wing contribution. The galaxy counts from the two sets of data were in good
agreement and agreed with the GAMA G10 data as well. This suggested that the galaxy
wings do not have an effect of more than 20\% on the faint galaxy fluxes used in \citet{driver}, 
and the effect is rather likely at $<10$\% level (Driver 2016, private communication).

As a simple test of consistency and as test of aperture size effects regarding faint galaxy 
photometry, we downloaded the HUDF frames\footnote{https://archive.stsci.edu/prepds/udf/}  and
ran our SExtractor photometry mentioned in Section 4.1 over the $B$--band image.  We did this 
in two ways, first with a standard setup using 2.5 for the Kron papameter, and then doubling it 
to 5.0, essentially doubling most elliptical aperture sizes.  Other parameters were fixed, 
the background being estimated in a 24 pixel thick annulus outside the apertures.  We do find, 
as e.g. \citet{Benitez} indicate, that many galaxies {\em close to the detection limit} become 
significantly brighter when using the larger apertures.  However, the systematic effect in our 
simple test is not nearly as large as 50\% for the light missed, but rather closer to 20\%.  Moreover, 
when summing up all the sources in the field over the whole detected magnitude range, the missing 
light fraction is only 6\%.

Thus, while it well may be that significant amounts of light are missed in 'typical' galaxy 
photometric techniques, it might be difficult to imagine that those surveys where specific care is 
taken against such effects could still miss more than, say, 20\% of the the total light because of 
the galaxy outskirts. Moreover, an obvious 
counter-argument for large fractions of missing light at the outskirts of galaxies is 
the absence of large SN populations in these areas. From the results of \citet{Bartunov}  
one finds that only $\sim$5--7\% of the total number of SN events occur at radii $r = (1.4-4)r_{25}$.  

\subsection{Low surface brightness galaxies}
Galaxies with very low surface brightness (LSBG) could escape detection altogether in galaxy counts, 
or their total brightness be significantly underestimated even if detected, due to various selection 
effects (e.g. \citealt{McGaugh}).  Hence it is in principle possible that these galaxies could 
contribute significantly to EBL while not adding to the integrated galaxy counts, even by a factor of 
two or more if suitable populations are constructed (e.g. \citealt{Vai}).  However, 
though such galaxies exist (e.g. Malin 1, the most famous case, \citealt{impey}), 
deep optical surveys (e.g. \citealt{driver05}) or using radio/21--cm HI surveys as pin-pointers
(e.g. \citealt{doyle,haynes}) over the past two decades have failed to detect significant 
populations of field LSBGs. In rich clusters numerous {LSBGs} or ultra diffuse galaxies have 
recently been detected (e.g. \citealt{davies15,dokkum,koda}) but their total contribution
to the IGL remains small.  The consensus appears to be that less than 20\% of additional
light is contributed by the LSBGs  to the IGL (e.g. \citealt{driver99,driver05}). 
  
Note however that \citet{disney} have recently argued that dim and/or dark galaxies might still 
be evading surveys if such galaxies were strongly clustered.

\subsection{Intra-cluster and intra--group light}
Intra--cluster (ICL) and intra--group light, sometimes collectively also referred to as Intra--Halo 
Light (IHL, \citealt{cooray}), is a well established component among cosmic light sources.
It originates from stars stripped off from galaxies in the cluster formation phase or in later interactions
between galaxies or, perhaps, also from stars formed {\em in situ} in the intra-cluster gas 
(see e.g. \citealt{kapferer}). 

Diffuse light between the galaxies was first noted  by \citet{zwicky} in deep photographs of the Coma Cluster. 
Wide-field CCD cameras have enabled deep surface 
brightness imaging of the intergalactic light in Coma (e.g. \citealt{adami05}),
Virgo (e.g. \citealt{mihos05, rudick}) and several other nearby 
and more distant clusters and galaxy groups (for a review see \citealt{mihos}).

The fraction of intra-cluster light of the total cluster luminosity varies between ~15--40\% 
for the nearby big clusters \citep{Ciardullo2004}.
For clusters in the redshift range $z\sim$0.2--0.4 some studies  \citep{Burke2015} find 
 that the ICL fractions are decreasing with redshift from  $\sim$25 to $\la$5\% 
while in others \citep{Guennou2012,Presotto2014}  the values are still high,
 $\sim$ 25\%, or even higher \citep{adami2016}. From six clusters at redshifts of $z\sim$0.8--1.2 
\citet{Burke2012} find ICL fractions of $\sim$1--5\%.
 
In galaxy groups the fraction of diffuse light varies even more strongly from undetectable in loose 
groups to $\sim$30--40\% in many compact groups (see e.g. \citealt{DaRocha05,white03}). 
 
The intra-cluster SN events account for $\la 20\%$ of the SN rate for the clusters, in agreement 
with the estimates of the diffuse light fraction \citep{Gal-Yam, graham}. 
For the intergalactic SN events in groups an upper limit of $\la 32$\% 
has been set by the fraction of apparently hostless Type Ia SNe \citep{mcgee}.

Although the diffuse light fraction in individual clusters and dense groups may be high, up to 40\%, it is 
important to remember that rich clusters contribute only a few percent of the total cosmic starlight while ~80\% of 
the light comes from individual field galaxies or loose groups, like the Local Group, with luminosities  
$< 10^{11}L_{\sun}$  \citep{eke}. Therefore, hardly more than 10\% is added to the integrated galaxy light 
by the contributions from ICL and intra-group starlight.

\subsection{Other sources of diffuse light}
Because of Lyman line and continuum absorption the redshift range of light sources contributing 
to sky brightness at $\lambda \la 450$ mm is limited to $z\la3.5$. 
Contributions by primordial objects such as population III stars or direct collapse black 
holes are thus excluded.

Decaying or annihilating dark matter candidate particles, such as neutrinos, WIMPs and axions, 
have been proposed as possible sources of diffuse background radiation fields. On the other side, 
the EBL might qualify as an important discovery channel for the elusive dark matter particles; 
for a review see \citet{overduin04,overduin08} and the update in \citet{henry}. Recently, 
\citet{cooray16} have discussed the possibility that the NIR background {\em fluctuations} could
partly originate from decaying axions with mass around 4 eV, located mainly in the halos of 
clusters and groups. None of the three particle species have been found to produce enough radiation 
to qualify as serious contributor to the {\em mean} intensity of the EBL in UV \citep{henry}, optical 
or NIR \citep{cooray16} domain. Furthermore, axions and WIMPS should be strongly concentrated
to the dark matter halos of clusters, groups and individual galaxies. Even if some part of the diffuse 
intra-cluster, intra-group or galaxy halo diffuse light were caused by decaying particles, 
instead of stars, this would not change the amount of diffuse light as determined by the observations.
Only a smoothly distributed diffuse light component, present also in the general field outside the 
clusters, could have escaped the observations.

\end{document}